\documentclass{aa}  

\usepackage{listings}
\usepackage{adjustbox}
\usepackage{supertabular}
\usepackage{layout}
\usepackage{multirow}
\usepackage{appendix}
\usepackage{bm}

\usepackage[T1]{fontenc}      % permet d'utiliser tous les caractères de votre clavier
\usepackage{hyperref}
\usepackage{amsmath}
\usepackage{amssymb}
\usepackage{mathrsfs}
\usepackage{animate}
\usepackage{graphicx}

\usepackage{subfigure}
\usepackage{enumitem}
\usepackage{systeme}
\usepackage{amsfonts}
\usepackage[makeroom]{cancel}

\usepackage{comment}

\usepackage{gensymb}

\usepackage{xcolor}
\usepackage{color}
\newcommand{\Br}{B_\mathrm{r}}
\newcommand{\Brs}{B_{\mathrm{r}}(r_\odot)}

\newcommand{\fK}{f_{\mathrm{K}}}
\newcommand{\fss}{f_{\mathrm{ss}}}

\newcommand{\M}{\mathrm{M}}
\newcommand{\omegasun}{\Omega_\mathrm{Sun}}

\newcommand{\ra}{r_\mathrm{A}}
\newcommand{\rs}{r_\odot}
\newcommand{\rss}{r_\mathrm{ss}}
\newcommand{\uphi}{u_\varphi}
\newcommand{\ua}{u_\mathrm{A}}
\newcommand{\ur}{u_\mathrm{r}}
\newcommand{\vwind}{v_\mathrm{wind}}

\newcommand{\re}{r_\mathrm{e}}
\newcommand{\sige}{\sigma_\mathrm{e}}

\newcommand{\rexp}{r_\mathrm{exp}}

\newcommand{\cs}{c_\mathrm{s}}

\newcommand{\game}{\gamma_\mathrm{e}}
\newcommand{\gamp}{\gamma_\mathrm{p}}
\newcommand{\gams}{\gamma_\mathrm{s}}

\renewcommand{\mp}{m_\mathrm{p}}

\newcommand{\np}{n_\mathrm{p}}

\newcommand{\Ps}{P_\mathrm{s}}
\newcommand{\rc}{r_\mathrm{c}}

\newcommand{\Teo}{T_\mathrm{e0}}
\newcommand{\Tp}{T_\mathrm{p}}
\newcommand{\Tpo}{T_\mathrm{p0}}
\newcommand{\Ts}{T_\mathrm{s}}
\newcommand{\Tso}{T_\mathrm{s0}} 
\newcommand{\To}{T_\mathrm{0}} 
\newcommand{\uc}{u_\mathrm{c}}

\newcommand{\xs}{x_\mathrm{s}}

\newcommand{\risoe}{r_\mathrm{iso|e}}
\newcommand{\risop}{r_\mathrm{iso|p}}
\newcommand{\risos}{r_\mathrm{iso|s}}

\newcommand{\ntil}{\Tilde{n}}

\newcommand{\ntils}{\Tilde{n}_\mathrm{s}} 
\newcommand{\sums}{\sum_\mathrm{s = \{p,e\}}} 
\newcommand{\phiref}{\phi_\mathrm{ss}}
\newcommand{\dang}{\delta_{\theta, \varphi} (\rs)}

\newcommand{\dt}{\delta t}

\newcommand{\Qang}{Q_\mathrm{ss}}
\newcommand{\ars}{\alpha_\mathrm{s}}
\newcommand{\arss}{\alpha_\mathrm{ss}}

\newcommand{\phio}{\phi_\odot}
\newcommand{\thetao}{\theta_\odot}
\newcommand{\phirss}{\phi_\mathrm{ss}}
\newcommand{\thetarss}{\theta_\mathrm{ss}}

\newcommand{\dphio}{\delta \phi_\odot}
\newcommand{\dthetao}{\delta \theta_\odot}
\usepackage{txfonts}
\begin{document} 

   \title{Testing the flux tube expansion factor - solar wind speed relation with Solar Orbiter data}

   %\subtitle{I. Overviewing the $\kappa$-mechanism}

   \author{J-B. Dakeyo  \inst{1,2} 
           \and A.P. Rouillard \inst{2} \and V. Réville\inst{2} \and P. Démoulin \inst{1,3} \and M. Maksimovic \inst{1} \and A. Chapiron \inst{2} \and R. F. Pinto \inst{2} \and  P. Louarn\inst{2} %\fnmsep\thanks{Just to show the usage
          %of the elements in the author field}
          }

   \institute{LESIA, Observatoire de Paris, Universit\'e PSL, CNRS, Sorbonne Universit\'e, Universit\'e de Paris, 5 place Jules Janssen, 92195 Meudon, France\\
              \email{jdakeyo@irap.omp.eu}
        \and
            IRAP, Observatoire Midi-Pyrénées, Universit\'e Toulouse III - Paul Sabatier, CNRS, 9 Avenue du Colonel Roche, 31400 Toulouse, France
         \and
             Laboratoire Cogitamus, 75005 Paris, France
             }

   \date{Received \today; accepted --}

% \abstract{}{}{}{}{} 
% 5 {} token are mandatory
 
  \abstract
   {The properties of the solar wind measured in situ in the heliosphere are largely controlled by energy deposition in the solar corona which in turn is closely related to the properties of the coronal magnetic field. Previous studies have shown that long duration and large scale magnetic structures show an inverse relation between the solar wind velocity measured in situ near 1 au and the expansion factor of the magnetic flux tubes in the solar atmosphere.
    }
   {The advent of the Solar Orbiter mission offers a new opportunity to analyse the relation between solar wind properties measured in situ in the inner heliosphere and the coronal magnetic field. We exploit this new data in conjunction with models of the coronal magnetic field and the solar wind to evaluate the flux expansion factor - speed relation.}
   {We use a Parker-like solar wind model, the "isopoly" model presented in previous works, to describe the motion of the solar wind plasma in the radial direction, and model the tangential plasma motion due to solar rotation with the Weber $\&$ Davis equations. Both radial and tangential velocities are used to compute the plasma trajectory and streamline from Solar Orbiter location sunward to the solar 'source surface' at $\rss$. We then employ a Potential Field Source Surface (PFSS) model to reconstruct the coronal magnetic field below $\rss$ to connect wind parcels mapped back to the %surface with the low corona/
   photosphere.
}
   {We find a statistically weak anti-correlation  between the in-situ bulk velocity and the coronal expansion factor, for about 1.5 years of solar data. %across all scales of magnetic structure. 
   Classification of the data by source latitude reveals different levels of anticorrelation, which is typically higher when Solar Orbiter magnetically connects to high latitude structures than when it connects to low latitude structures. We show the existence of fast solar wind that originates in strong magnetic field regions at low latitudes and undergoes large expansion factor. We provide evidence that such winds become supersonic during the super radial expansion (below $\rss$), and are theoretically governed by a positive correlation v-f. We find that faster winds on average have a flux tube expansion at a larger radius than slower winds.} 
   {An anticorrelation between solar wind speed and expansion factor is present for solar winds originating in high latitude structures in solar minimum activity, typically associated with coronal hole-like structures, 
   but this cannot be generalized to lower latitude sources. 
   We have found extended time intervals of fast solar wind associated with both large expansion factors and strong photospheric magnetic fields. Therefore, the value of the expansion factor alone cannot be used to predict the solar wind speed. Other parameters, such as the height at which the expansion gradient is the strongest must also be taken into account. }


\maketitle
%
%-------------------------------------------------------------------

%%%%%%%%%%%%%%%%%%%%%%%%%

%%%%%%%%%%%%%%%%%%%%%%%%%%%%%%%%%%%%%%%%%%%%%%%%%%%%%%%%
\section{Introduction} \label{sec:intro}
%%%%%%%%%%%%%%%%%%%%%%%%%%%%%%%%%%%%%%%%%%%%%%%%%%%%%%%%

%%%%%%%%%%%%%%%%%%%%%%%%%

% In-situ and source properties correlation
Decades of heliospheric exploration have attempted to understand the relation between solar wind properties measured in-situ and its source in the solar corona \citep{wang_sheeley1990, neugebauer1998, abbo2016, bemporad2017}. These properties are defined during the formation of the wind and can evolve greatly during propagation in the interplanetary medium. Measurements taken closer to the wind sources with Solar Orbiter (SolO) and Parker Solar Probe (PSP) allow to alleviate some of the propagation effects and improve our comparative studies of wind properties with coronal structures. 

A large fraction of the open magnetic field lines along which the solar wind originates are %typically
rooted in coronal holes clearly visible in ultraviolet solar imaging. These holes only occupy a fraction of the solar photosphere. %\citep{bohlin1977}. 
At solar minimum, coronal holes persist at high latitudes near the poles and cover about 15-20\% of the total solar surface area \citep{bohlin1977}. These polar holes can extend down to a latitude of 60$^\circ$ in each hemisphere \citep{wang1996, Wang2010}. As solar activity increases, the area of polar holes shrinks to less than 5\% of the total surface area near solar maximum and, in contrast to the rather persistent polar holes, these low-latitude holes tend to evolve rapidly in response to solar activity \citep[see, e.g.,][]{broussard1978, Insley1995, dorotovic1996}. % \citep{broussard1978, Insley1995}. 
This cyclic evolution induces strong topological changes in the magnetic fields and therefore the solar wind properties.  

% Expansion factor introduction
Since open field lines only occupy a small fraction of the total solar surface, magnetic equilibrium will force solar wind flux tubes to expand faster than predicted by a simple spherical expansion \citep{zirker1977}. This super radial expansion is described by the expansion factor $f(r)$ typically defined as: 
\begin{equation}
f(r) = \frac{ \Br(\rs)}{\Br(r)} \frac{\rs^2}{r^2}
\end{equation}
where $\Br(r)$ is the magnetic field radial component at the radial position $r$ and $\rs$ is the Sun radius. Using a magnetic model of the solar corona and an empirical extrapolation of the 1 au velocity $v$ from the $f(\rss)$ value computed at the source surface radius $\rss$, \citet{wang_sheeley1990} noticed an anticorrelation pattern between the 1 au solar wind velocity $v$ and $f(\rss)$ for observations on large spatial and temporal scales. 
Their analysis used a magnetostatic reconstruction of the solar atmosphere called the Potential Field Source Surface (PFSS) model where the magnetic field is supposed to be potential within $\rs <r<\rss$ and radial for $r>\rss$. Using a similar approach, \citet{arge2000} found a better anti-correlation by also considering the distance of the wind source to the closest coronal hole boundary. Other magnetic connectivity studies, and MHD simulation studies have looked at different solar sources and their correlation with in-situ properties, but they either considered short time intervals, or a given type of magnetic structure \citep{riley2012, pinto2016, reville2017, wallace2020, badman2023, yardley2024}. In addition, the MHD simulation study of \citet{pinto2016} highlights that discrepancies of the global v-f anticorrelation can exist, supporting that the global scaling law of \citet{arge2000} requires adjustments. 
Thus, we cannot say to what extent the v-f anti-correlation can be generalized to all observed solar wind time periods. 

% Magnetic connectivity definition
Magnetic connectivity mapping is a central step in these studies. 
The tracing of a spacecraft magnetic connectivity to the photosphere has typically been divided in two spatial domains. Starting from the spacecraft, a magnetic field line is first traced through the interplanetary medium to the upper corona by usually following the nominal 'Parker spiral'. In a second step, a model of the complex coronal magnetic field is usually considered to trace magnetic field lines from the upper corona down to the photosphere.

For the first step, a recent study by \citet{dakeyo2024} compared the different existing mapping methods. While the ballistic backmapping approach (constant wind speed) is the most commonly used, they showed that the best practice remains to consider wind acceleration and corotation effects, especially when studying different wind speed streams. 
For the second step,
past research comparing different coronal models with the PFSS reconstructions conclude that PFSS provides good results particularly near solar minimum when the large-scale currents are negligible and the source surface can be considered near spherical \citep{Riley2006, arden2014}. The PFSS model uses as input the radial component of the photospheric magnetic field $\Brs$ (provided in the form of photospheric magnetograms). The distribution of $\Brs$  control the topology of the coronal magnetic field and determine the trajectory followed by magnetic field lines of interest, so the path of the young solar wind as the plasma is frozen in the magnetic field.

% Insity SW properties
Complementary observational constraints can be obtained with in-situ measurements. The radial evolution of the thermodynamic properties of the solar wind has been studied by many authors \citep{Schwartz1983radial, Hellinger2011heating, Hellinger2013proton, Stverak2015electron, SanchezDiaz2016, SanchezDiaz2019, maksimovic2020} using the large coverage of heliocentric distances provided by the Helios missions \citep{1981_ref_helios}. In particular, the study of \citet{maksimovic2020} has revealed that the slow wind pursues its acceleration at large radial distances (0.3 to 1 au), outside of the main wind acceleration region ($\lesssim 20~\rs$). Other studies using more recent data from the PSP mission \citep{fox2016solar} conclude on similar trends \citep{halekas2022, dakeyo2022}.  

Thus, the radial evolution of the velocity beyond the main acceleration region should be accounted for in the solar wind models and should have some influence on our estimates of magnetic field connectivity. In most connectivity studies, the velocity used to model the streamline spiral is assumed to be purely radial and constant with radial distance \citep[e.g.][]{snyder1966, krieger1973, burkholder2019, rouillard2020, Badman2020, gritton2021}.
However, in practice this is not the case with respect to studies of the radial evolution of the observed wind speed \citep{maksimovic2020, halekas2022, dakeyo2022}.

% Explanation of the isopoly equation
In order to model the radial evolution of the solar wind speed for different wind types, we have recently developed a simple "isopoly" fluid model \citep{dakeyo2022}. Derived from a two-fluid Parker-type solar wind model, this approach assumes two different thermal regimes with radial distance: an isothermal corona, representing the region where coronal heating is effective, and a polytropic expansion in the solar wind. This simple model allows some significant wind acceleration near the Sun (below 15 $\rs$), while matching in-situ measurements of the radial velocity, temperature and density profiles recorded beyond the solar corona. However, this modeling does not take into account the near-Sun magnetic topology associated with the super radial expansion.

% Probematic
With all the above concerns in mind, we would like to explore in this article to what extent the v-f relation could be generalized to different types of solar wind. %a non-exhaustive and long-time description of the solar wind, including in-situ constraints on the radial and tangential evolution of the wind. 
The context of the Solar Orbiter mission is particularly interesting for a deeper analysis of solar wind properties with radial distance and to improve on current connectivity models \citep{muller2020}.

In Sect.~\ref{sec:connectivity_methods} we aim to perform a statistical study using magnetic connectivity and solar wind modeling. Then, we infer the source location of a large set of in-situ measurements made by Solar Orbiter. Given the limitations of existing techniques and models, we have chosen to perform a fast calculation of the near-Sun magnetic topology based on PFSS extrapolation, a refined streamline calculation based on the Parker spiral including isopoly description (radial acceleration) and \citet{weber_davis1967} tangential flow (corotational effects). In Sect.~\ref{sec:Results}, we present the magnetic connectivity results on the global v-f induced relation, and compare them to the previous results. % the similarities with coronal holes characteristics. 
In Sect.~\ref{sec_expansion_factor_and_wind_speed}, we provide theoretical justification to explain the results. 
Finally, in Sect.~\ref{sec:conclusion}, we summarize the main results of our study and discuss their implications for in-situ to source relation and global solar wind modeling.

%%%%%%%%%%%%%%%%%%%%%%%%%

%%%%%%%%%%%%%%%%%%%%%%%%%%%%%%%%%%%%%%%%%%%%%%%%%%%%%%%%
\section{Connectivity method and modeling} \label{sec:connectivity_methods}
%%%%%%%%%%%%%%%%%%%%%%%%%%%%%%%%%%%%%%%%%%%%%%%%%%%%%%%%

%%%%%%%%%%%%%%%%%%%%%%%%%

The magnetic connectivity of a wind streamline is computed as a two-step process, including interplanetary magnetic field modeling and near-Sun magnetic field modeling. 
We include acceleration and corotational effects in the wind description. Indeed, both can modify the final mapped longitude in the solar corona.

%%%%%%%%%%%%%%%%%%%%%%%%%%%%%%%%%%%%%%%%%%%%%%%%
\subsection{Radial evolution: IsoPoly equations}
\label{sec:isopoly_equations}
%%%%%%%%%%%%%%%%%%%%%%%%%%%%%%%%%%%%%%%%%%%%%%%%

% Existing solar wind modeling

Our aim is to model the solar wind evolution with heliocentric distance using in situ measurements from SolO as the upper boundary conditions. Since the model should be run on every available data point of the statistical study, a requirement of our model is to run fast. The "isopoly" fluid model proposed by \citet{dakeyo2022} fulfills these requirements with short computational time. Moreover, it has already provided a successful description of the solar wind observations of Helios and PSP between 0.1 au and 1 au. 
The conservation of momentum is 
\begin{equation}
   n\, \mp\, \ur \frac{d \ur}{d r} = - \sums \frac{d \Ps}{d r}  - n\, \mp \frac{G \, M}{r^2},
  \label{eq_momentum_sans_hypothese}
\end{equation}
with $n$ the density, $\Ps$ the plasma pressure, $\mp$ the proton mass, $M$ the Sun mass, and where the sum over the species $s$ is taken over protons ($\mathrm p$)  and electrons ($\mathrm e$). The temperature is
\begin{align}
      \Ts(r) = \Tso \: \bigg( \frac{n(r)}{n(\risos)} \bigg)^{\gams -1} 
      \rightarrow
    \left\{
        \begin{array}{ll}
            \text{if} \quad r\: \leq \: \risos : \quad \gamma_s = 1   \\
            \text{if} \quad r\: > \: \risos : \quad  \gamma_s > 1 
        \end{array}
    \right.,
    \label{eq_Ts_isopoly}
\end{align}
where $\risos$ is the distance below which the expansion is isothermal, and $\gams$ is the polytropic index. The density can be expressed using mass flux conservation, where $n \: u \: r^2 = C$, and $C$ is a constant determined from observations. For further information on isopoly equations, we refer to \citet{dakeyo2022} and \citet{chen2022}.

%%%%%%%%%%%%%%%%%%%%%%%%%%%%%%%%%%%%%%%%%%%%%%%%
\subsection{Tangential evolution: Weber-Davis equations}
\label{sec:weber_davis_equations}
%%%%%%%%%%%%%%%%%%%%%%%%%%%%%%%%%%%%%%%%%%%%%%%%

% Streamline modeling , spirals
In the solar corona, the solar wind initially corotates with the Sun in the inertial frame. Beyond a certain distance the plasma moves mostly radially at supersonic speeds. The relative rotation between the released plasma and its source creates the Parker spiral pattern \citep{parker1958}, which is an idealised representation of the magnetic field lines in the interplanetary medium. It corresponds to the velocity streamlines based on the magnetic field lines viewed in the corotating frame (also called the Carrington frame). Throughout this manuscript, the trajectories of the plasma parcels viewed in the solar corotating frame are referred to as streamlines.

% Limitation ballistic backmapping
In most connectivity studies, the velocity used to model the streamline spiral is assumed to be purely radial and constant with radial distance \citep[e.g.][]{snyder1966, krieger1973,SanchezDiaz2016, rouillard2020, Badman2020, gritton2021}.
However, in practice this is not the case as shown in a number of recent studies \citep{maksimovic2020, dakeyo2022, halekas2022}. Moreover, even a partial corotation of a solar wind parcel with its source induces a tangential flow $\uphi$, tightening the streamline and changing the calculated field line trajectory. 

\citet{weber_davis1967} carried out a study showing for an MHD solar wind outflow that the plasma is almost in quasi-rigid rotation with the Sun very low in the corona ($\uphi = \omegasun\, r$). The corotation becomes weaker with distance $r$, especially above the Alfvén critical point $\ra$, and then tends asymptotically to a non-corotating flow far from the Sun ($\uphi =0$). This modeling is supported by other recent studies \citep{macneil2022, koukras2022}.
% Introduction of the tangential speed and corotating distance
Based on a given radial speed profile $\ur(r)$, \citet{weber_davis1967} derived the following expression for the tangential speed $\uphi(r)$:

\begin{equation}
    \uphi(r) = \frac{\omegasun \: r}{\ua(\ra)} \:
    \frac{\ua(\ra) - \ur(r)}{ 1 - M_A(r)^2 },
    \label{eq_tangential_speed}
\end{equation}
where $\ua(r) = |B_r(r)| /\sqrt{\mu_0 \: \rho(r)} $ is the Alfv\'en speed profile, $\rho(r)$ the total mass density, and $M_A=\ur(r)/\ua(r)$ the Alfvén mach number.

% Magnetic field typical values
To compute $\ur(r)$ we use the five radial isopoly speed profiles of \citet{dakeyo2022} which are interpolated in order to match the Solar Orbiter in-situ bulk speed measurements. From these interpolated profiles we can compute the associated Alfv\'en speed $\ua(r)$ if we also know the radial evolution of $B_r$ provided by Solar Orbiter (detailed later in Sect.~\ref{subsec:Solar_data_treatment}). Using these values and the isopoly density profiles, we compute the Alfv\'en speed $\ua(r)$. Then, $\uphi(r)$ is computed with Eq.~\eqref{eq_tangential_speed}.

% Tangential speed in the Parler spiral equation
Incorporating $\uphi$ in the calculation of the local streamline longitude $\phi$ based on the expression of the Parker spiral, is achieved by subtracting the relative angular speed of the plasma to $\omegasun$ \citep{macneil2022}: 
\begin{align}
    \phi(r) = \phiref + \int_{\rss}^{r} \frac{ \omegasun - \uphi(r') / r' }{\ur(r')} dr' 
    \label{eq_parker_spiral_with_corot}
\end{align}
where $\phiref$ is the longitude location at $\rss$, and $r$ the distance from which is computed the backmapping.
For more details about the tangential speed profiles associated with the isopoly radial solutions, please refer to \citet{dakeyo2024}.

%%%%%%%%%%%%%%%%%%%%%%%%%%%%%%%%%%%%%%%%%%%%%%%%%%%%%%%%
\subsection{Solar Orbiter data set and treatment} \label{subsec:Solar_data_treatment}
%%%%%%%%%%%%%%%%%%%%%%%%%%%%%%%%%%%%%%%%%%%%%%%%%%%%%%%%

% Want to establish of data set for SolO data
The three instruments that make the Solar Wind Analyzer (SWA) suite \citep{owen_SWA2020} are the Proton-Alpha particle Sensor (PAS), the Electron Analyzer System (EAS), and the Heavy Ion Sensor (HIS). For the magnetic field data, the magnetometer (MAG) provides 3D measurements of the interplanetary magnetic field \citep{Horbury_MAG2020}. Based on the currently available data, we restrict our dataset to the instruments PAS, HIS, and MAG.
% How we process with data 
The Solar Orbiter observations we use for the study cover from 01/08/2020 to 17/03/2022. This includes data up to the end of the solar minimum activity period. Since we are not studying short timescales here, we have calculated average values of the observations over 30 minutes to smooth the fast variations and optimize the connectivity process (computation time). 

%%%%%%%%%%%%%%%% no simultaneous data and ICME removal %%%%%%%%%%%%%%%%%%%%%%%

Time intervals for which all the instruments do not provide simultaneous data were not considered and we also removed ICMEs from our Solar Orbiter dataset using the criteria of \citet{Elliott2012}. We discard the measurements for which at least one of the two following criteria on the plasma $\beta$ and the proton temperature $\Tp$ is satisfied~: 
$(i)$ $\beta < 0.1$,   
$(ii)$ $\Tp / T_{ex} < 0.5$, 
where $T_{ex}$ is a temperature predicted by the scaling law $T_{ex}~=~486.5 \times u - 1.2476\times 10^5K$ established by \citet{Lopez1986solar}, for which $\Tp$ is rescaled with solar distance by the solar wind predicted temperature $T_{ex}$. We consider that ICMEs have a duration of at least 6 hours. In addition to these criteria, we remove the 24 h before and 15 h after each detected ICME. Furthermore, we consider wind measurements faster than 800 km/s as potential ICMEs, and therefore also remove them.

\begin{figure*}[t]
    \centering
    \includegraphics[width = 18.cm]{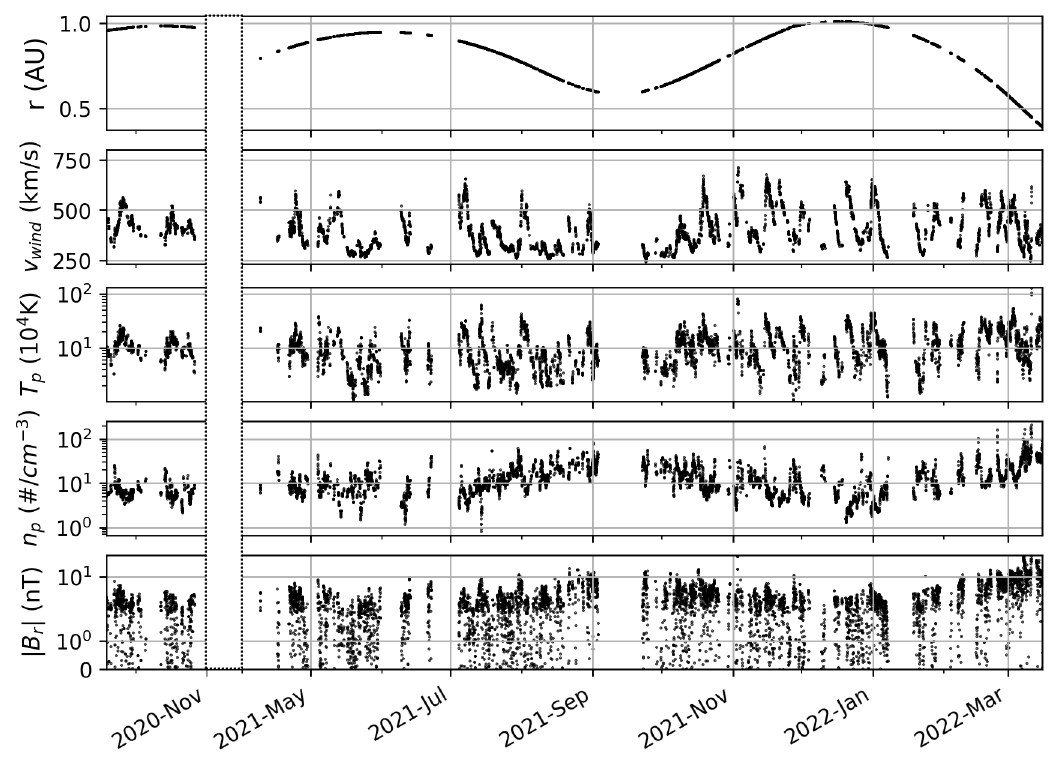}
    \caption{Time series of Solar Orbiter measurements from 01/08/2020 to 17/03/2022 in the period of minimal solar activity. The top panel shows the radial position of SolO. The bulk speed $\vwind$, proton temperature $\Tp$ and proton density $\np$ are provided by the instrument PAS, and the absolute value of the magnetic field radial component $\Br$ by the instrument MAG. The time intervals for which both instruments do not provide simultaneous data, and the observations not associated to a complete backmapping process, are not displayed. Each data point is a 30 minutes average. }
    \label{fig_solo_data_time_series}
\end{figure*}

The synchronized and filtered data from Solar Orbiter are shown in Fig.~\ref{fig_solo_data_time_series}. There is a significant data gap between November 2020 and April 2021 because no synchronized data was available for PAS and MAG. This reduces the real observable time to an equivalent of 1 year of continuous data. However, the wind speed sampling is good enough to assume that the observation depicts a global picture of the solar wind characteristics (apart from an under sampling of the faster winds due to the minimal solar activity).

%%%%%%%%%%%%%%%%%%%%%%%%%%%%%%%%%%%%%%%%%%%%%%%%%%%%%%%%
\subsection{The PFSS model and magnetograms:} \label{subsec:magnetogram_PFSS}
%%%%%%%%%%%%%%%%%%%%%%%%%%%%%%%%%%%%%%%%%%%%%%%%%%%%%%%%

% PFSS algorithm advantages and limitation
The most widely used coronal model is the PFSS, mainly because of its ease of use and short computation time. Above an assumed source surface $\rss$ typically placed between 1.5 and 5 $\rs$ (most commonly 2.5 $\rs$), all field lines are assumed to be open to the interplanetary medium. To determine the trajectory of a given plasma parcel below $\rss$, it is necessary to compute the coronal magnetic equilibrium %around the Sun's surface 
to determine the open and closed field lines as well as the complex connectivity. The reconstruction is quite accurate for studying the solar corona near solar minimum, but it neglects the electric current and assumes the magnetic field to be fully potential, and is therefore less accurate during active solar periods \citep{Riley2006}.
Moreover, the PFSS model forces the source surface to be fixed at the same height for all magnetic structures, and while the estimation of $\rss = 2.5~\rs$ is typically considered as the best one \citep{arden2014}, coronographic observations have shown that all magnetic structures have not the same typical opening height \citep{sheeley1997, wang1998}. Algorithms have been developed recently to determine the optimal source-surface height through a direct comparison of PFSS calculations with white-light imaging \citep{Poirier_2021}. Considering that our study focuses on a relatively quiet solar period (minimal solar activity), PFSS can still be considered as an appropriate modeling method.

Regarding second order limitation of the PFSS, a study by \citet{rouillard2016} suggest that PFSS tends to overestimate $\fss$ near the HCS compared to MHD modeling. Indeed, although closed field lines reconstructed near the HCS embed large magnetic gradients for both modeling methods, the PFSS magnetic reconstruction compute strongly diverging $\Br$ field lines near the HCS. This leads to a strong $\Br$ decrease with very large $\fss$ values when PFSS and MHD are compared at the same height. Moreover, the region around the HCS mentioned above with large $\fss$ could be extended up to four times larger by PFSS than by MHD \citep{rouillard2016}. This could lead to an overestimation of the $\fss$ value of the magnetic field lines surrounding the HCS. These discrepancies have not been more deeply quantified in the literature, but \citet{Riley2006} and \citet{rouillard2016}, both comparing MHD with PFSS, agree on a global coherence between the two methods in minimum solar activity in terms of global magnetic topology.

% ADAPT Magnetogram description
Our input to the PFSS model are ADAPT-GONG magnetograms \citep{adapt_ref2013} \footnote{\href{https://gong.nso.edu/adapt/maps/gong/}{https://gong.nso.edu/adapt/maps/gong/}}. These magnetograms are produced by continually assimilating new observations and by also applying a flux-transport model to simulate the poleward migration of magnetic elements during the solar cycle. There are 12 different ADAPT magnetogram realizations produced every two hours. Considering the complexity of the full mapping process over such an important dataset, and the relatively close similarity of all 12 realizations of PFSS calculations \citep{Li2021}, we only consider the last realization of the ADAPT maps, since they are the only ones which remain available on the ADAPT GONG web site for the studied time period.  %

% PFSS details
Our PFSS algorithm uses a spherical harmonic decomposition of the magnetogram \citep{schatten1969}. The order of spherical harmonics $l$ is set to $l=20$. This constitutes an intermediate resolution compared with the resolution of the magnetogram, allowing to capture the complexity of the corona on the small and large scales while limiting artificial artifacts due to the decomposition itself \citep{poduval2004,Toth2011}. Higher $l$ could be considered to refine the mapping accuracy, but for the purpose of the present statistical study, we aim to keep a reasonable computational time and focus on global tendencies.

\begin{figure*}[t]
    \centering
    \hspace{-.5cm}
    \includegraphics[width = 18.5cm]{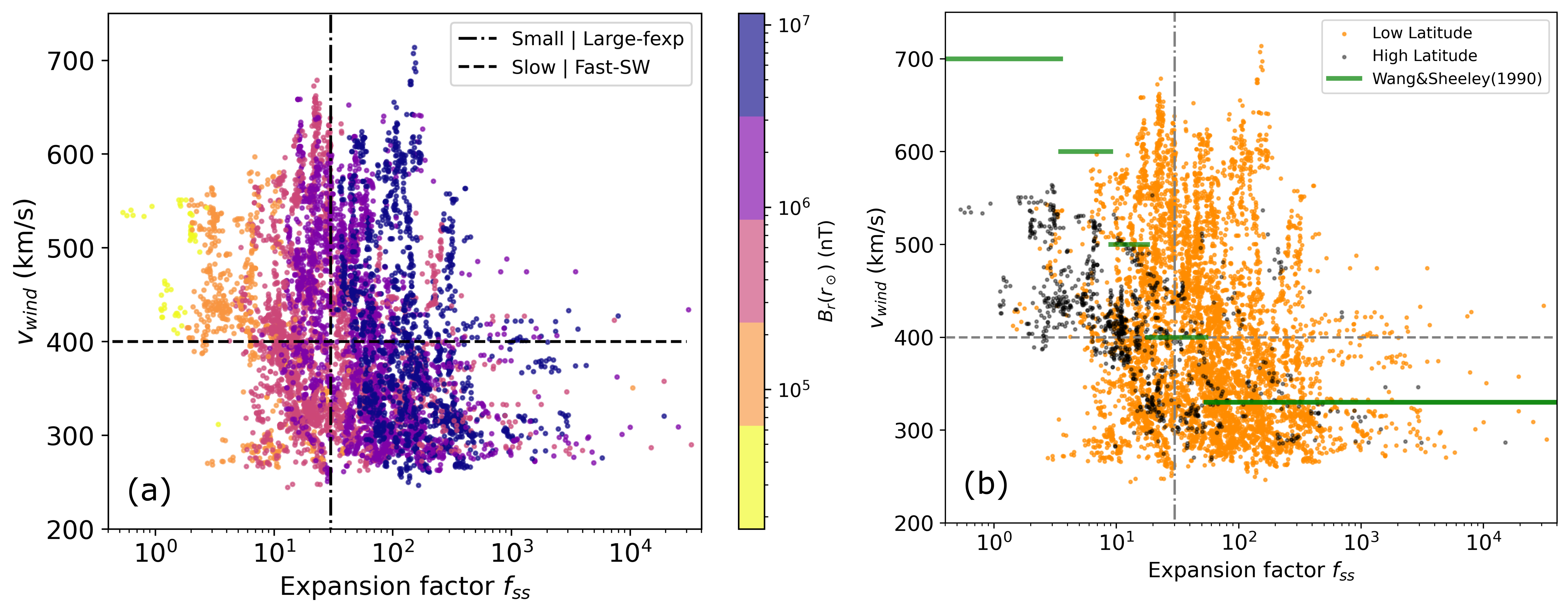}
    \caption{Relationship between the measured velocity $\vwind$ and the final expansion factor value $\fss$ computed with PFSS from the back-mapping applied to the data in the Figure \ref{fig_solo_data_time_series}. Panel (a): the wind speed $\vwind$ measured by Solar Orbiter as a function of the expansion factor $\fss$ 
    calculated using PFSS at the source surface (located at $\rss$). The mapping results cover from 01/08/2020 to 17/03/2022. The %Sun's surface 
    photospheric magnetic field is displayed with the color coding shown in the color bar, and with partial transparency to limit the masking effect. The distribution has a Pearson correlation coefficient of -0.27. 
    Panel (b): Same as the panel (a) but colored in black for low unsigned latitudes (< 45\degree) and in orange for high latitude footpoints (> 45\degree) for each mapped observation. The typical range of values from \citet{wang_sheeley1990} study, mapping observation at 1 au, are displayed by the horizontal green bars. The high and low latitudes distributions have a Pearson correlation coefficient of -0.51 and -0.24 respectively.
    }
    \label{fig_u_fexp_B0_magneto_low_high_latitude}
\end{figure*}

%%%%%%%%%%%%%%%%%%%%%%%%%

%%%%%%%%%%%%%%%%%%%%%%%%%%%%%%%%%%%%%%%%%%%%%%%%%%%%%%%%
\section{Statistical results on v-f correlation} \label{sec:Results}
%%%%%%%%%%%%%%%%%%%%%%%%%%%%%%%%%%%%%%%%%%%%%%%%%%%%%%%%

% No clear anti-corraltion v-f
We use the mapping technique, described in Sect.~\ref{sec:connectivity_methods}, to study the relation between the solar wind speed $\vwind$ measured in-situ at Solar Orbiter and the expansion factor $\fss = f(\rss)$ found at the source surface. 
The results are shown in Fig.~\ref{fig_u_fexp_B0_magneto_low_high_latitude}. The panel (a) presents the solar wind speed as a function of the expansion factor, i.e. the v-f relation, for all data points. The color code is defined in terms of the intensity of the radial magnetic field component at the photospheric footpoint. 
Squaring the plot in small and large values of $\fss$ and $v$, the smaller and larger $\fss$ values ($\lesssim$ 6 and $\gtrsim$ 300) are mainly restricted respectively to moderately fast and slow solar wind. 
This is what could be expected from the inverse relation reported by \citet{wang_sheeley1990} and \citet{arge2000}. We notice however that the anti-correlation is unclear for a large fraction of the measured wind. In fact, the overall v-f distribution has a Pearson and Spearman rank correlation coefficients of -0.27 and -0.29 respectively, which is not high enough to suggest a global v-f anticorrelation.

% Comparison to previous similar backmapping study
Moreover, period of fast wind streams ($>$ 500 km/s) measured by Solar Orbiter that map back to high values of $\fss$ and $|\Brs|$ are of particular interest to the present study (top right square area). These are puzzling as they are not the common results found in the known sources of high-speed streams such as coronal holes.

% Relation fss and Br0
Regarding magnetic field and expansion factor relation, we see that magnetic flux tubes with high $\fss$ values have strong photospheric field strengths $|\Brs|$ (Fig.~\ref{fig_u_fexp_B0_magneto_low_high_latitude}, panel (a)). This result was also reported in \citet{wang2009} and is coherent with the fact that, during increasing solar activity, magnetic field lines with strong expansion factors tend to be rooted at low latitudes in the active region belt. 

Next, we notice that the lowest $|\Brs|$ values are only associated with $\vwind>$~400 km/s). More globally, $\fss$ is related to $|\Brs|$ (we checked that this is not due to the plotting point ordering, so a masking effect of earlier plotted points, by plotting $|\Brs|$ in function of $\fss$).

% Split v-f plot in latitude and find anti-correlation 
In order to disentangle the differences on the v-f relation between our study and the previous ones, we classify the data points according to their estimated source latitudes and split the data between high ($>45^\circ$) and low ($<45^\circ$) unsigned latitude of the source regions. The classification results are shown in the panel (b) of Fig.~\ref{fig_u_fexp_B0_magneto_low_high_latitude}. The plasma originating from high latitudes (black dots) follow an anti-correlation qualitatively similar as the one presented by \citet{wang_sheeley1990}, while low latitude sources do not present a specific correlation (orange dots). These classification results are supported quantitatively by the fact that the Pearson and Spearman rank correlation coefficients of the v-f distribution are -0.51 and -0.59, respectively, for the high latitude sources, while only -0.24 and -0.26, respectively, are obtained for the low latitude sources. 
This suggests that winds observed originating at low latitudes present an expansion factor value which is not intrinsically related to solar wind asymptotic speed in solar minimal activity and that other factors could come into play.

For comparison, the v-f relation obtained by \citet{wang_sheeley1990} is shown as green horizontal solid bars. Some of the mapped high-latitude structures are qualitatively consistent with this v-f relation, but we must note that the two studies differ considerably in terms of the connectivity process. In fact, the relation found by \citet{wang_sheeley1990} is based on  daily measurements near-1 au, %solar wind measurements 
averaged over a 3 month sliding window.   
The sliding average window acts as a temporal filter, removing short duration structures. The connectivity operated by \citet{wang_sheeley1990} is a direct projection of the Earth's Carrington coordinates onto the Sun, using a default Sun-Earth transit time of 5 days. No consideration is given to magnetic field lines tracing, wind speed, or variation of the travel time. In addition, the magnetogram low resolution involves a spatial filtering, so with only large-scale magnetic structures remaining. All these settings imply that the method of \citet{wang_sheeley1990} is highlighting large-scale structures with long time duration, which match with large coronal holes characteristics. Considering a minimal solar activity, such source characteristics are typically found at high latitude.

\begin{figure}[h]
    %\centering
    \hspace{-.3cm}
    \includegraphics[width = 9.3cm]{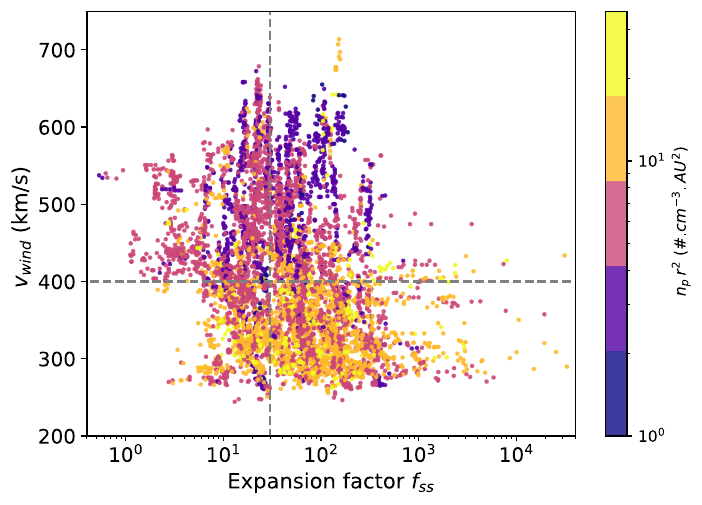}
    \caption{Same as Figure~\ref{fig_u_fexp_B0_magneto_low_high_latitude} but colored by measured in-situ density corrected by $r^2$.
    }
    \label{fig_u_fexp_n_r2}
\end{figure}

% Density observations 
Figure~\ref{fig_u_fexp_n_r2} is similar to panel (a) of Figure~\ref{fig_u_fexp_B0_magneto_low_high_latitude}, but colored with the measured density corrected by the radial distance $r^2$. 
We observe that slow winds ($< 450$ km/s) are generally denser than fast winds ($> 450$ km/s), which is coherent with previous solar wind studies \citep{Schwartz1983radial, maksimovic2020, dakeyo2022}.  
We also notice that fast winds with $\fss \approx 100$ are less dense than fast winds originating from low expansion regions. These weaker solar wind densities show that the significant magnetic field expansion is not compensated by enhancements in plasma escape from the subsonic corona. 

% Uncertainties of the mapping
The streamline tracing from the probe to the Sun is subject to deviation from several effects such as corotating interaction regions (CIRs) and uncertainties on bulk speed estimation (as well as a more precise account of corotational effects). 
However, these effects are difficult to include in a statistical way into the mapping process. To account for their influence, we have recomputed the results of Fig.~\ref{fig_u_fexp_B0_magneto_low_high_latitude}, applying a perturbation on $\phi(r)$ of $\pm$5\degree and $\pm$10\degree at $\rss$. The results are presented in the Appendix~\ref{appendix:backmapping_uncertainties}. We observe in Fig.~\ref{fig_u_fexp_B0_magneto_pm5_10} that the overall shape presented in Fig.~\ref{fig_u_fexp_B0_magneto_low_high_latitude} is similar for the $\pm$5\degree and $\pm$10\degree panels.
This indicates low variability in global trends. 
Consequently, this supports the existence of all the different mapped wind populations, and the reliability of the statistical mapping process itself. Please refer to the Appendix~\ref{appendix:backmapping_uncertainties} for further details.

%%%%%%%%%%%%%%%%%%%%%%%%%

%%%%%%%%%%%%%%%%%%%%%%%%%%%%%%%%%%%%%%%%%%%%%%%%%%%%%%%%
\section{Expansion factor and asymptotic wind speed } \label{sec_expansion_factor_and_wind_speed}
%%%%%%%%%%%%%%%%%%%%%%%%%%%%%%%%%%%%%%%%%%%%%%%%%%%%%%%%

%%%%%%%%%%%%%%%%%%%%%%%%%

% Existence of fast wind with large f
The results of the back-mapping study highlight the fact that, although fast solar wind streams originating from large $\fss \gtrsim 50$ regions are unexpected, they represent a non-negligible fraction of the wind measured in the interplanetary medium.

%%%%%%%%%%%%%%%%%%%%%%%%%%%%%%%%%%%%%%%%%%%%%%%%
\subsection{Solar wind equations with super expansion} \label{sec_equations_super_expansion}
%%%%%%%%%%%%%%%%%%%%%%%%%%%%%%%%%%%%%%%%%%%%%%%%

% Existence of fast wind with large $\fss$ and theoretical justification
We therefore consider what the physical implications of fast wind with large $\fss$ might be, and how such a wind could be explained theoretically. The fast solar wind acceleration processes have been related to the $\fss$ value by \citet{wang1993}, who introduced the idea that $\fss$ and the efficiency of the heating mechanism are anti-correlated. In fact, they show that a wind originating from open magnetic field lines with small $\fss$ values experiences sustained heating over a higher range of altitudes including above the sonic point inducing greater terminal speeds to be reached. In contrast, solar wind forming along field lines undergoing large expansion is heated primarily below the sonic point, thereby increasing plasma density at the expense of efficient plasma acceleration. This point of view is greatly supported and used in the literature \citep{verdini2007, wang2009, chandran2011, pinto_rouillard2017, shi2023}. At first glance this theory does not explain how large $\fss$ region would host high-speed streams.

% Hugoniot equation justification for fSW with large fss
Flows in a diverging flux tube can be described by the Hugoniot equation, which is more commonly used in fluid mechanics in the de Laval nozzle \citep{seifret1947}. Extending this modeling to the solar wind leads to the equations developed by \citet{kopp_holzer1976}. 
In their momentum Eq.(6), the Mach number gradient $d\M/dr$ multiplied by the factor $(\M^2-1)$, and the flux tube area gradient $d\mathrm{A}/dr$ are of same sign in the case of a subsonic flow regime ($\M <1$) and of opposite sign for a supersonic flow regime ($\M >1$). This implies that if the plasma speed of the wind is large enough low in the solar corona, a super radial expansion can lead to an increase of the flow speed. 
Figure 4 of \citet{kopp_holzer1976} illustrates the transition between the two types of flow regime which happens for a critical value of maximal expansion factor value $f_m$ (depending on the other model parameters). Below this limit, the flow  
is completely subsonic up to several $\rs$ ($r \approx 4.5 \rs$ in their example), and above the limit, the sonic point location jumps much closer to the Sun ($r \approx 1.3 \rs$), giving supersonic flow and large speeds in the high corona. 
Moreover, the larger $f_m$, the more the wind is accelerated.
Since the $f_m$ critical value depends on the modeled coronal condition (temperature profiles, polytropic index value), a single value representative of all wind cases shown in Figure \ref{fig_u_fexp_B0_magneto_low_high_latitude} cannot be determined.

\begin{figure*}[ht]
    \hspace{-.3cm}
    \includegraphics[width = 18.5cm]{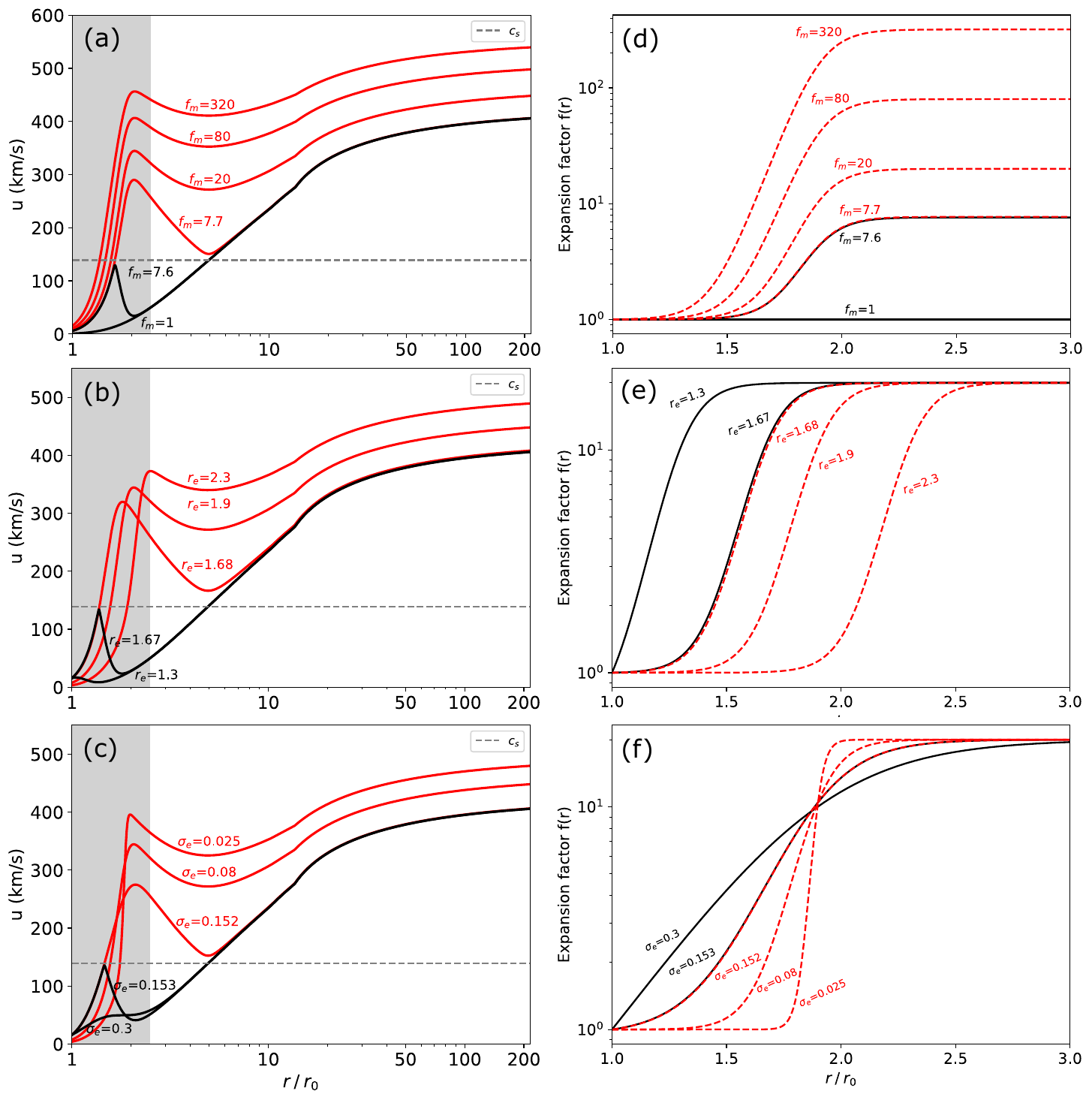}
    \caption{ Isopoly models with super-radial expansion described by Eq.~\eqref{eq_fK}. The solutions embedding subsonic and supersonic regimes below $2.5~\rs$ are respectively displayed in black solid line and red solid lines. The left panels show isopoly solutions for the same initial parameters ($\Tpo=1.63 \text{ MK}, \:\Teo=0.71\text{ MK},\: \risop=13.6\rs, \: \risoe=10.3\rs, \: \gamp=1.52, \: \game=1.23$), with varying expansion factor parameters ($f_m, \re, \sige$), and the right panels show the associated $f(r)$ profiles; 
    Panel (a): Different values of $f_m$ (maximum expansion factor obtained for large $r$); 
    Panel (b): Different values of $\re$ (radius at which the super expansion almost stops); 
    Panel (c): Different values of $\sige$ (broadness of the expansion region). 
    The $f(r)$ profiles associated to panels (a), (b) and (c) are displayed on panels (d), (e), and (f) respectively.
    For all the $f(r)$ parameters not displayed on the panel are set with ($f_m=20$, $\re=1.9 \, \rs$, $\sige=0.08 \, \rs$). 
    While PFSS is not used here, we still mark the region located below the source surface with the gray area as a guide for comparison. 
    }
    \label{fig_profil_isopoly_fexp_example}
\end{figure*}

Following the development of \citet{kopp_holzer1976} applied to the isopoly hypotheses, we complete the isopoly equations including the super-radial expansion with an expansion factor profile $f(r)$. The conservation of mass flux is re-written as follows~:
\begin{align}
    n \: u \: f \: r^2  = C,
\end{align}
where $C$ a constant determined from observations. 
The resolution of Eq.~\eqref{eq_momentum_sans_hypothese} including $f(r)$ follows the same development as in \citet{dakeyo2022}, with an additional term in the derivative of the density: 
\begin{align}
   \frac{d \ntils}{d r} = - \frac{1}{n(\risos)} \frac{C}{f r^2} \bigg[ \frac{2}{ur} + \frac{1}{u^2} \frac{du}{dr} + \frac{1}{uf} \frac{df}{dr} \bigg]
\end{align}

where $\ntil = n(r)/n(\risos)$.
Its inclusion in the momentum Eq.\eqref{eq_momentum_sans_hypothese} leads to:
\begin{align}
    \frac{du}{dr} 
     \underbrace{ \bigg[ 1 -  \frac{c^2}{u^2} \bigg] }_{a(r,u)} 
     = \underbrace{  \frac{1}{u r} \bigg[  c^2 \:  
        %\bigg(2 - f r \: \frac{d (1/f)}{dr} \bigg) 
        \bigg(1 + \frac{r}{2} \: \frac{d \log f}{dr} \bigg)
        - \frac{G \, M}{r} \bigg] }_{b(r,u)} 
\label{eq_momentum_all_term_detailled_fexp}
\end{align}
where $c^2=\sums \cs^2 \xs$, $\: \xs= \ntil^{\: \gams -1} $, and from Eq.~\eqref{eq_Ts_isopoly} the distance $\risos$ below which the expansion is isothermal. 
Equation~\eqref{eq_momentum_all_term_detailled_fexp} is similar to the isopoly Equation (B8) presented in \citet{dakeyo2022}, with an extra term related to $f(r)$ expressing the effect of the super radial expansion on the change in velocity. 
This extra term is positive for a flux tube monotonously expanding outward.

% Existence two solution sub-supersonic above rss
The starting point of the resolution is fixed at the critical radius $r_c$, 
where both $a(r,u) =0$ and $b(r,u) =0$.  The first condition implies that this always occurs for $u=c$ independently of the expansion profile $f(r)$. 
The second condition, $b(r,u) =0$, defines $r_c$.
In \citet{dakeyo2022}, $r_c$ was found to be present always in the isothermal region of the model. From Eq~\eqref{eq_momentum_all_term_detailled_fexp} a larger tube expansion should imply a lower $r_c$, then $r_c$ stays in the isothermal region, and we remind $\xs= 1$ for $\gams=1$.
Finally, the transonic solution crosses the sonic point at $r_c$ with the condition $du/dr~\neq~0$. In contrast to the resolution of Eq.~\eqref{eq_momentum_sans_hypothese} (equivalent to the case $f(r)=1$) where the sonic point is unique, solving Eq.~\eqref{eq_momentum_all_term_detailled_fexp} with $f(r)$ presents possibly two distinct sonic points with only one related to a solar wind solution. The determination of the appropriate $\rc$ value is given numerically by first integrating the equation from the largest $\rc$ possible value (satisfying $b(r,u)=0$), with  positive $u(r)$ derivative, in the sunward direction.  %between the possible ones. 
If the computed solution preserves $u(r<\rc) < \uc$, it is kept. Otherwise, we remake the calculation with the other possible $\rc$ value (the smallest one),  with positive $u(r)$ derivative at $\rc$.  

%%%%%%%%%%%%%%%%%%%%%%%%%%%%%%%%%%%%%%%%%%%%%%%%
\subsection{Fast wind with large $\fss$ values} \label{subsec_fast_wind_large_f}
%%%%%%%%%%%%%%%%%%%%%%%%%%%%%%%%%%%%%%%%%%%%%%%%

An example of the resolution of Eq.~\eqref{eq_momentum_all_term_detailled_fexp} is presented in Fig.~\ref{fig_profil_isopoly_fexp_example}, using as $f(r)$ the widely-used profile of \citet{kopp_holzer1976}~:

\begin{equation}
\centering
     \fK (r) = \frac{f_m \: e^{(r- \re)/\sige} + f_1}{ e^{(r- \re)/\sige} + 1}  
     \label{eq_fK}
\end{equation}

where the parameter $f_1$ is selected to set $\fK (\rs) = 1$ as follows:
\begin{equation}
\centering
    f_1 = 1 - (f_m -1) \: e^{(\rs - \re)/\sige}
\end{equation}

The expansion profile $\fK (r)$ represents a monotonously increasing super radial expansion with $r$, with the main extra expansion concentrated just below $r=\re$ with a broadness $\sige$. The expansion at large distance ($r>>\re+\sige$) is defined by $f_m$.

% Effect of sub-supersonic expansion on the modeling
Figure~\ref{fig_profil_isopoly_fexp_example} extends the parametric study of the velocity profile dependence presented by \citet{kopp_holzer1976} to the isopoly model, adding the influence of the other free parameters of Eq.~\eqref{eq_fK}. 
Each left panel of Figure~\ref{fig_profil_isopoly_fexp_example} shows the variation for only one of the three expansion factor parameters; $f_m$ for panel (a), $\re$ for panel (b) and $\sige$ for panel (c). Each right panel shows the expansion factor profiles associated with the curves in the corresponding left panel. This example is computed with the intermediate wind speed, for the population \textbf{C} (population referred from \textbf{A} to \textbf{E} for slow to fast wind respectively) of \citet{dakeyo2022}, using the same input parameters ($\Ts$, $\gams$, $\risos$) for all curves. The input values can be found in Table 1 of \citet{dakeyo2022} and are summarized in the caption of Figure~\ref{fig_profil_isopoly_fexp_example}. 
 
%%%%%%%%%%%%%%%%%%%%%%%%
% f_m influence
The value of $f_m$ sets the amplitude of the super-radial expansion as shown in panel (a) of Figure~\ref{fig_profil_isopoly_fexp_example}. Its effect is similar to that of a multiplicative factor (although not directly proportional to $f(r)$ in Eq. \ref{eq_fK}). Since the interplanetary magnetic field is almost uniform, one can directly relate $f_m$ to the magnetic intensity $B_{r,0}$ at the base of the flux tube.
%r_exp influence
The parameter $\re$, shown on panel (b), sets the radius at which the super expansion almost stops. Considering that the super radial expansion depends on the global magnetic equilibrium and the typical height of closed surrounding magnetic structures, $\re$ is directly related to the height of the magnetic structure being considered next to the flux tube path. 

The parameter $\sige$, displayed on panel (c), sets the typical distance over which the super-radial expansion occurs, i.e. the expansion region length. The super expansion occurs typically on a distance of the order of $\sim 5 ~\sige$. 
Nevertheless, as seen in the different curves for decreasing $\sige$ values, modifying only $\sige$ while keeping constant the other parameters also changes the effective expansion radius. Consequently its effect is dual on the $f(r)$ profile.

% Introduction of f-subsonic , f-super
In our study, we will refer to subsonic solutions below $\rss$ as "f-subsonic" and to the supersonic ones as "f-supersonic" solutions. The f-subsonic and f-supersonic solutions are shown as black and red solid lines, respectively, all along the article.

%Panel (a) 
Figure~(\ref{fig_profil_isopoly_fexp_example}a) shows similar results to \citet{kopp_holzer1976} for isopoly solutions with a limit value of $f_m$ at which there is a change from f-subsonic to f-supersonic solutions. Increasing the super expansion until $f_m=320$ allows to gain $\sim$150 km/s of speed, obtaining a 550 km/s wind at 1 au. Referring to the initial 5 isopoly populations \citep{dakeyo2022}, this leads at large distances ($r\geq 10\, \rs$) to a fast solar wind similar to \textbf{D} population (while the thermal plasma parameters are from \textbf{C} population).

%Panel (b)
Varying the parameter $\re$, Fig.~(\ref{fig_profil_isopoly_fexp_example}b) shows the same change of solution from f-subsonic to f-supersonic for larger $\re$ values. This indicates that the further the super expansion occurs, the more efficient it is to accelerate the wind. While Eq.~\eqref{eq_fK} sets no limit on $\re$, in practice, it is unrealistic to set $\re$ to values larger than $\rss \approx 2.5\, \rs$. Since PFSS results limit the flux-tube expansion below $\rss$ (Sect. \ref{subsec:magnetogram_PFSS}), we limit $\re$ to $2.3\, \rs$. With this limit and a moderate expansion ($f_m=20$), Fig.~(\ref{fig_profil_isopoly_fexp_example}b) indicates a possible extra speed at 1 au of $\sim$80 km/s.

%Panel (c)
A similar transition of solution is found for the last parameter $\sige$.  Figure~(\ref{fig_profil_isopoly_fexp_example}c) shows that f-supersonic solutions are modeled for low $\sige$ values. As discussed above in the same section, a change in $\sige$ value modifies both the expansion length and the expansion radius, so here one cannot determine to what extent the effective expansion length plays an important role in the f-subsonic and f-supersonic modeling. Nevertheless, in practice, lower values of $\sige$ could lead to a faster wind modeled by an f-supersonic solution. The gain of velocity is comparable to the results presented for $f_m$ and $\re$.

\begin{figure*}[t]
    \centering
    \hspace{-.5cm}
    \includegraphics[width = 18.5cm]{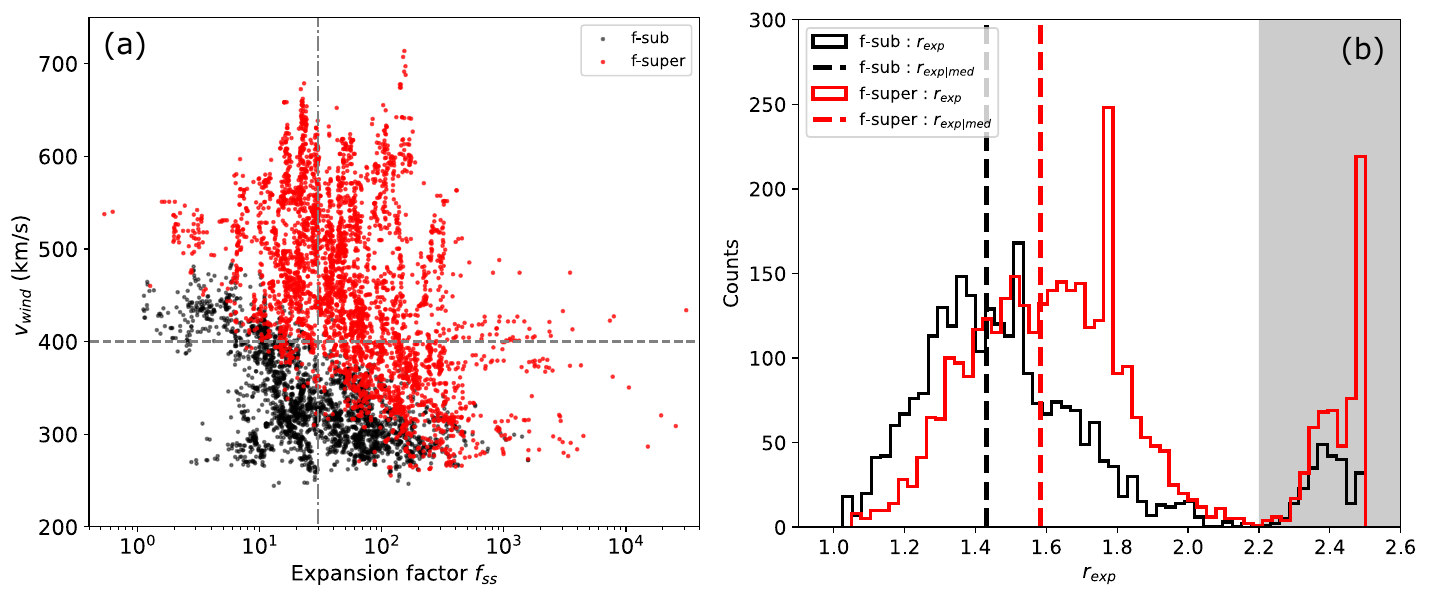}
    \caption{Wind speed and typical expansion properties of the f-subsonic and f-supersonic solutions. Panel (a): Same as panel (a) of Fig.~\ref{fig_u_fexp_B0_magneto_low_high_latitude} but colored (black or red) by the type of identified solutions (f-subsonic or f-supersonic, respectively) after computing isopoly models including $f(r)$ from PFSS. The isopoly parameters are interpolated from the 5 populations isopoly parameters of \citet{dakeyo2022}. The modeled observations include 42\% and 58\% of f-subsonic and f-supersonic solutions, respectively. Panel (b): Typical expansion radius from Eq.\eqref{eq_def_rexp} of the data presented on panel (a) classified by solution type. The f-subsonic and f-supersonic solutions are represented in black and red respectively. }
\label{fig_u_fexp_obs_supersonic_hexp}
\end{figure*}

Combining the effects of the three parameter variations presented in Fig.~\ref{fig_profil_isopoly_fexp_example}, the case in which the f-supersonic solutions provides the most acceleration is for a large global expansion ($f_m$ large), set at large expansion radius ($\re$ large) and on a narrow radial range (small $\sige$). This parametric study points out that for the same input parameters ($\Tpo, \Teo, \risop, \risoe, \gamp, \game$), the f-supersonic solution could model a faster wind speed on the order of 100 km/s larger.

% Effects on mapping results
Moreover, the f-subsonic and f-supersonic solutions embed different radial evolution of the speed. Indeed, the mapping results presented in Sect.~\ref{sec:Results} are based on f-subsonic solutions for all the mapped data. The f-supersonic solutions might imply alterations of the back-mapping result since the velocity profile is modified (so the transit time as well as the streamline local angle $\phi (r)$).
Based on the parametric study results (Fig.~\ref{fig_profil_isopoly_fexp_example}), the use of f-supersonic models, instead of the f-subsonic used initially, could tighten the streamline (higher velocity reached closer to the Sun), shorten the travel time and modify the isopoly estimated coronal parameters depending on the associated $f(r)$ profiles obtained from PFSS.

% Uncertainties due to mapping with f-supersonic solution
The change of the estimated isopoly coronal temperature will be discussed next in the Sect.~\ref{subsec_updated_isopoly_speed_profiles_and_param}. Regarding the streamline and travel time modification due to the use of f-supersonic solution, we have estimated to what extent the longitude computation at $\rss$ and the travel time are affected by comparing f-subsonic to f-supersonic backmapping computation. The details of the results are available in Appendix~\ref{appendix:backmapping_uncertainties_f_super_f_sub}. Finally, we have estimated that the streamline deviations are small enough in comparison to the backmapping process uncertainties presented in Appendix~\ref{appendix:backmapping_uncertainties_unquantified_devia}. Moreover, the travel time discrepancies do not significantly influence, in general, the magnetogram chosen to compute the magnetic topology at the wind time departure. Thus, accounting for f-supersonic modeling in the mapping process should not significantly affect the overall results of the study.

%%%%%%%%%%%%%%%%%%%%%%%%%%%%%%%%%%%%%%%%%%%%%%%% 
\subsection{Solar wind f-subsonic and f-supersonic expansion deduced from magnetograms} 
%\subsection{Solar wind model with expansion deduced from magnetograms} 
\label{subsec_flow_regime_expansion_height}
%%%%%%%%%%%%%%%%%%%%%%%%%%%%%%%%%%%%%%%%%%%%%%%%

We have shown in Sect.~\ref{subsec_fast_wind_large_f} that other parameters, such as super-expansion gradient and expansion radius, influence the final wind speed (without changing the coronal temperatures). 
This knowledge of modeling super expansion flow regimes can be subsequently included in the isopoly modeling used for mapping. In fact, keeping the same input parameters ($\Tpo, \Teo, \risop, \risoe, \gamp, \game$) and considering the $f(r)$ profiles obtained with PFSS, it is possible to determine if the resolution of the isopoly leads to a f-supersonic model instead of the standard f-subsonic initially used. One may question whether observations related to f-supersonic isopoly solutions may contain peculiarities associated with their $f(r)$ profile compared to f-subsonic models.

To compute isopoly profiles of mapped data with expansion modeling, we use the $f(r)$ profiles from the PFSS calculations for $r< \rss$, and we set $f(r)= f(\rss)$ for $r> \rss$ (the $\fss$ value depends of the field line). The isopoly parameters ($\Tpo, \Teo, \risop, \risoe, \gamp, \game$) from \citet{dakeyo2022} are interpolated using measured bulk velocity with those of the five isopoly populations. 
Solutions are computed with Eq.~\eqref{eq_momentum_all_term_detailled_fexp}. The computation of either a f-subsonic or a f-supersonic solution is not an arbitrary preset, and it is fully constrained by both the input parameters and $f(r)$, respecting the solving conditions mentioned at the end of Sect.~\ref{sec_equations_super_expansion}. So depending on the $f(r)$ profile, the same given input parameters may lead to two possible $\rc$ values, for which the f-subsonic solution is associated with the farthest critical radius while the f-supersonic solution is associated with the closest one (as explained in Sect.~\ref{sec_equations_super_expansion}).

Panel (a) of Fig.~\ref{fig_u_fexp_obs_supersonic_hexp} is similar to panel (a) of Fig.~\ref{fig_u_fexp_B0_magneto_low_high_latitude} showing the v-f relationship of the mapped data, but with the data classified by the type of solutions~: f-subsonic (in black) or f-supersonic (in red).  
The f-subsonic solutions are present for less than half of the mapped observations. %They also show a v-f anticorrelation. 
Moreover, the fast solar wind is only modeled by f-supersonic solutions. This implies that the role of super radial expansion in wind acceleration is of primary importance in modeling the asymptotic wind speed.

One may question the differences in expansion factor profiles between f-subsonic and f-supersonic solutions. To investigate this, we divided the mapped data into several $\fss$ bins and examined the shape of $f(r)$. 
Considering the typical $\fss$ values related to the different existing magnetic structures, we split the data with the boundary bin values $\fss~=~[7, 20, 50, 100, 250]$. Figure~\ref{fig_fprof_median_of_vf_plot_from_PFSS_fbins} presents the classified $f(r)$ profiles. The color code is the same as panel (a) of Figure 
\ref{fig_u_fexp_obs_supersonic_hexp}. 
The median profiles are plotted in cyan solid lines and dotted lines for f-subsonic and f-supersonic, respectively. The main result is that, for all the $\fss$ studied values, the $f(r)$ profiles associated with f-supersonic isopoly solutions, embed a later expansion radius than those associated with f-subsonic solutions. There is no significant difference in the expansion gradient of the $f(r)$ median profiles between the f-subsonic and f-supersonic solutions.

To quantify the difference in expansion radius between the two types of solution, we compute a generalized typical expansion radius $\rexp$, similar to $\re$, but computed for a general $f(r)$ profile. The expansion radius is defined as the radius at which occurs the main inflection point of $f(r)$ for an expansion increasing outward (i.e. the location of the zero of the second order derivative, while the first derivative is positive)~:
  \begin{align}
    \rexp = r \quad \text{where} \quad f^{''}(r) = 0, \quad \text{and} \quad f^{'}(r) > 0.
    \label{eq_def_rexp}
  \end{align}
In case $f^{''}(r) = 0$ is not verified at any radius for a given profile, we assume the inflection point has not been reached yet and then we set $\rexp = \rss$.

The expansion radius distributions are shown in panel (b) of the Figure~\ref{fig_u_fexp_obs_supersonic_hexp} with the same color code as in panel (a). The $\rexp$ distribution of the f-supersonic solutions is shifted to a higher altitude compared to the f-subsonic one, supporting the expansion radius results of Figure~\ref{fig_fprof_median_of_vf_plot_from_PFSS_fbins}. Considering that the expansion radius is related to the maximum altitude of the surrounding closed magnetic structures, the f-supersonic solutions originate from magnetic regions surrounded on average by higher, so typically horizontally more extended, magnetic structures. 

\begin{figure*}[t]
    \hspace{-.35cm}
    \includegraphics[width = 18.5cm]{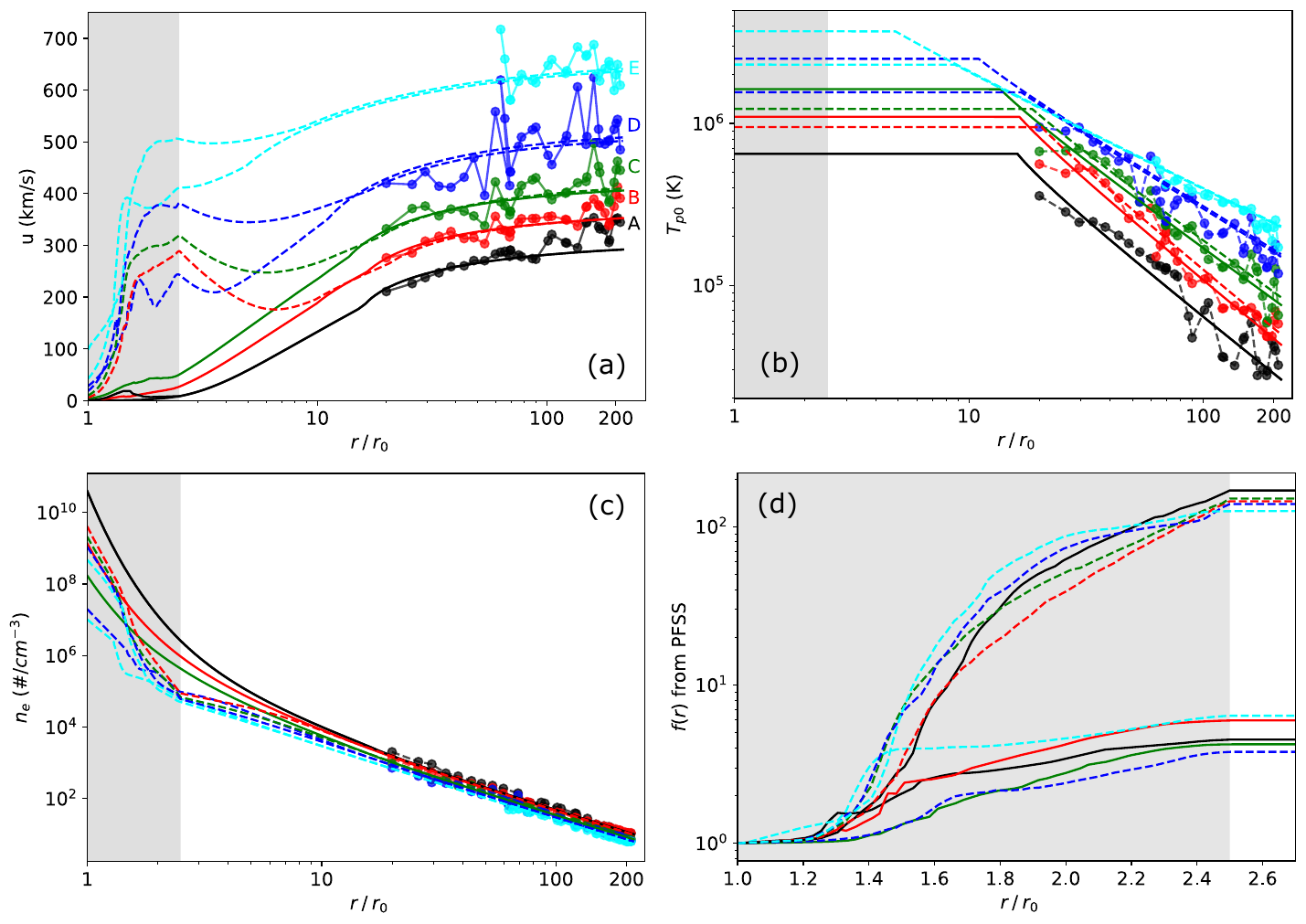}
    \caption{Updated isopoly models from Eq.~\eqref{eq_momentum_all_term_detailled_fexp} associated with expansion factor profiles computed from PFSS, fitted to the data set used by \citet{dakeyo2022}. The f-subsonic and the f-supersonic solutions are plotted in solid and dashed lines respectively. Five isopoly models are computed for the two bins $\fss < 7$ and $ 100 < \fss <250$. Panel (a)~: Updated velocity profiles; Panel (b)~: Updated proton temperature profiles; Panel (c)~: Updated density profiles; Panel (d): Corresponding median $f(r)$ profiles calculated from the $f(r)$ profile obtained by the PFSS reconstruction. The data used for fitting are added in panels (a), (b), and (c) as dots linked with straight segments. The region of super radial expansion (up to $\rss=2.5~\rs$) is delineated by the gray shaded area. The number of expansion profiles used to compute the median profiles of the [\textbf{A, B, C, D, E}] wind populations in panel (d) are [14, 16, 123, 235, 4] and [89, 360, 326, 148, 103], for the $\fss < 7$ and $ 100 < \fss <250$, respectively. }
\label{fig_u(r)_reajut_f-subsup_to_PSP_helios_data}
\end{figure*}

We notice that some values of $\rexp$ are identified close to $\rss=2.5 \: \rs$ ($\rexp> 2.2 \: \rs$) and they form a small secondary distribution. This could be due to PFSS limitations implying two artifacts. First, the inflection point could be not reached below $2.5 \: \rs$.
Second, when the associated field lines are close to the Heliospheric Current Sheet (HCS), the source surface imposes a local divergent $f(r)$ variation. More precisely, an artificial divergence of the field lines is implied by the source surface at the top of the closed magnetic field. 
This indicates that the source surface height could be too low, at best, or more generally that the PFSS modeling is not sufficient for these regions near the HCS. In summary, the $\rexp$ values above $2.2 \: \rs$ are an artifact due one of the two above limitations of PFSS modeling. 
However, they do not change the overall interpretation since they are relatively marginal cases. Moreover, the median values computed excluding $\rexp >2.2~\rs$ (dotted vertical lines in Figure~\ref{fig_u_fexp_obs_supersonic_hexp}) show similar $\rexp$ separation, with an offset of $\sim 0.15 \:\rs$ between the f-subsonic and f-supersonic $\rexp$ distributions.

\begin{table*}[ht]
\centering
\scalebox{0.95}{
%\hspace{-1.7cm}
\begin{tabular}{cccccc}
    %%%%%%%%%%%%%%%%%%%%%%%%%%%%%%%%%%%%%%%%%%%%%%%%%%%
    \hline %\hline 
    Wind type & A & B & C & D & E    
     \tabularnewline
    \hline 
    %%%%%%%%%%%%%%%%%%%%%%%%%%%%%%%%%%%%%%%%%%%%%%%%%%%
    $\Tpo$ (MK) & 0.65 &  1.1 - \textbf{0.95} & 1.63 - \textbf{1.23} & 2.51 - \textbf{1.56} & 3.71 - \textbf{2.31}
     \tabularnewline
    %\hline 
    %%%%%%%%%%%%%%%%%%%%%%%%%%%%%%%%%%%%%%%%%%%%%%%%%%%
    $\risop$ ($\rs$) & 16.1 & 16.4 - 16.9 & 13.6 - 19.9 & 9.2 - 15.2 &  2.9 - 8.9
     \tabularnewline
    %\hline 
    %%%%%%%%%%%%%%%%%%%%%%%%%%%%%%%%%%%%%%%%%%%%%%%%%%%
    $\To$ (MK) & 0.72 & 0.96 - \textbf{0.88} & 1.17 - \textbf{0.97} &  1.63 - \textbf{1.16}& 2.3 - \textbf{1.6}
     \tabularnewline
    \hline 
    %%%%%%%%%%%%%%%%%%%%%%%%%%%%%%%%%%%%%%%%%%%%%%%%%%%
    $u_{0}$ (km/s) & 0.02 - 1 & 1 - 6 & 3 - 10 & 28 - 18 & 98 - 42
     \tabularnewline
    %\hline  
    %%%%%%%%%%%%%%%%%%%%%%%%%%%%%%%%%%%%%%%%%%%%%%%%%%%
    $u_{1au}$ (km/s) & 292 & 350 - 370 & 406 - 416 & 492 - 527 & 623 - 645 
     \tabularnewline
    \hline  
    %%%%%%%%%%%%%%%%%%%%%%%%%%%%%%%%%%%%%%%%%%%%%%%%%%%
    $n_{0}$ ($10^{7} \#/cm^3$) & 3960 & 136 - 409 & 17 - 210 & 2 - 109 & 1 - 48
     \tabularnewline
    %\hline  
    %%%%%%%%%%%%%%%%%%%%%%%%%%%%%%%%%%%%%%%%%%%%%%%%%%%
    $n_{1au}$ ($\#/cm^3$) & 10 & 9.7 & 7.2 & 6.2 & 5.4
     \tabularnewline
    %\hline  
    %%%%%%%%%%%%%%%%%%%%%%%%%%%%%%%%%%%%%%%%%%%%%%%%%%%
    \hline
\end{tabular}  
}
\caption{Updated isopoly input parameters ($\Tpo, \risop$) associated to the isopoly curves accounting for expansion factor modeling in Figure~\ref{fig_u(r)_reajut_f-subsup_to_PSP_helios_data} (two top lines). The estimated mean coronal temperature $\To = \frac{1}{2} (\Tpo + \Teo)$ is shown on the third line.  The associated coronal and at 1 au velocities ($u_0$, $u_{1au}$), and densities ($n_0$, $n_{1au}$), respectively, are shown in the four bottom lines. The initial parameters from \citet{dakeyo2022} have been modified in order to fit the in-situ measured  temperatures and velocities of their 5 wind populations with the computed wind solutions including $f(r)$ modeling. The 1 au density values are also calibrated to the 5 wind populations in order to have a direct comparison with \citet{dakeyo2022} isopoly models. Bold values are features of primary interest. }
\label{tab_updated_isopoly_param}
\end{table*}

%%%%%%%%%%%%%%%%%%%%%%%%%%%%%%%%%%%%%%%%%%%%%%%%
\subsection{Updated isopoly profiles and parameters} 
\label{subsec_updated_isopoly_speed_profiles_and_param}
%%%%%%%%%%%%%%%%%%%%%%%%%%%%%%%%%%%%%%%%%%%%%%%%

% Motivation to update isopoly profiles
The influence of the expansion factor on the wind models presented in Sect.~\ref{sec_expansion_factor_and_wind_speed} has revealed that if the wind populations are modeled by f-supersonic solutions, the isopoly models presented in \citet{dakeyo2022} may embed a modification of their speed and density profile, and of their coronal temperature values $(\Tpo, \Teo)$. Indeed, accounting for f-supersonic modeling could both decrease the required isopoly coronal temperatures and change the velocity profile in the near Sun region ($r \leq 10 \: \rs$). 

% Double acceleration effect for converging then diverging flux tube
Moreover, it should be noted the presence of some non monotonic $f(r)$ profiles below $\rss$ as shown %on panel (a) and (b) of 
in Figure~\ref{fig_fprof_median_of_vf_plot_from_PFSS_fbins}. They are analyzed in Appendix~\ref{appendix:f(r)_profiles_details}. For such profiles, $f(r)$ first decreases, then $f(r)$ increases with distance, which could favor wind acceleration by both subsonic and supersonic flow regimes. Indeed, a subsonic wind in the low corona in a converging flux tube receives an additional acceleration according to the de Laval nozzle effect. Next, if this wind accelerates enough to overcome the local sound speed, the wind could undergo further acceleration from the diverging part of the flux tube. The flow acceleration is then twofold, thus providing a way to model fast winds with f-supersonic solutions and a strong influence of $f(r)$, even if $f(\rss)$ is not large. %considering small super radial expansion.

% Aim to include $f(r)$ in isopoly equations wih $f(r)$ and update params
In order to take into account the above concerns, we update the isopoly speed profiles and their input parameters from \citet{dakeyo2022} in accordance with the influence of $f(r)$. We use the $f(r)$ profiles computed from the PFSS algorithm presented in Figure~\ref{fig_fprof_median_of_vf_plot_from_PFSS_fbins}. 
% Modification of the protons temperature
Since f-supersonic solutions are mostly associated to intermediate and fast winds, as seen in panel (a) of Fig. \ref{fig_u_fexp_obs_supersonic_hexp}, and that these wind populations are mainly driven by protons \citep{dakeyo2022}, we only updated the proton coronal temperatures $\Tpo$ and left $\Teo$ unchanged.

% Choice of the $f(r)$ profiles from PFSS
Given the variety of $f(r)$ profile shapes obtained from PFSS, we illustrate only isopoly models for small and large super radial expansions. To do so, we use the profiles classified by $\fss$ bins presented in Figure~\ref{fig_fprof_median_of_vf_plot_from_PFSS_fbins}, %with bins limits $[7,20,50,100,250]$, 
and we keep the profiles in the first bin, $\fss <7$, and in the second to last bin, $100 < \fss <250$. For each of these two $\fss$ bins, the profiles are further classified according to their measured speed, i.e. we assign each mapped observation to one of the 5 isopoly wind populations. This results in a total of 10 subgroups of $f(r)$ profiles. We keep the median profiles of each subset. Consequently, for each $\fss$ bin, all isopoly populations have a unique median $f(r)$ profile that best matches the observed speed to which they correspond. We must note that the $f(r)$ median profiles are sensitive to the statistics used to compute each of them, so the updated isopoly profiles represent an estimate of what the speed profiles and their isopoly input parameters could be based on PFSS modeling, not a unique representation of the wind population behavior. 

% Updated isopoly profile description. Generality
Figure~\ref{fig_u(r)_reajut_f-subsup_to_PSP_helios_data} shows all the updated isopoly profiles, with the velocity on the panel (a), the proton temperature on the panel (b), the density on the panel (c) and the associated $f(r)$ medians profiles on the panel (d). The updated isopoly parameters are presented in the Table \ref{tab_updated_isopoly_param}, where the left value corresponds to the bin $\fss<7$, the right value to the bin $100 < \fss < 250$, and in case the parameter is unchanged it remains a single value. 

% Updated isopoly profile description. Results
The wind population \textbf{A} (in black) is exclusively modeled by f-subsonic solutions, and present a deceleration region between 1.5 $\rs$ and 2.5 $\rs$ (panel (a) of Figure~\ref{fig_u(r)_reajut_f-subsup_to_PSP_helios_data}). 
The wind populations \textbf{B} and \textbf{C} (red and green, respectively) are associated to both f-subsonic and f-supersonic models depending on the $f(r)$ profile selected. The f-supersonic solutions show a deceleration region above 2.5 $\rs$ until $\sim~5-7~\rs$. 
Regarding the fast wind populations \textbf{D} and \textbf{E} (dark and light blue, respectively), they are exclusively modeled by f-supersonic solutions.  

The panel (b) of Figure~\ref{fig_u(r)_reajut_f-subsup_to_PSP_helios_data} presents updated isopoly temperature profiles that are similar to the ones computed in \citet{dakeyo2022}. However, in accordance with Table \ref{tab_updated_isopoly_param}, a main difference is that $\Tpo$ is lower especially for the faster winds. 

Next, the panel (c) shows that updated density profiles are significantly affected by super radial expansion below $\rss$ in the case of f-supersonic modeling. Indeed, for the same wind asymptotic speed (red and green curves), the coronal densities are smaller with f-supersonic modeling compared to the f-subsonic one in a broad region around $\rss$, while the effect is inverse in the low coronal region. For fast winds (blue colors), the coronal density of the f-supersonic solutions has also the same behavior with a sharp decrease between $\sim 1.5$ and 2 $\rs$. This sharp decrease induces a strong outward pressure gradient which is at the origin of the sharp wind acceleration of f-supersonic solutions. 
In summary, the main effect of $f(r)$ is to modify the plasma density via the mass flux conservation. This implies enhanced pressure gradients in both the larger and more localized flux tube expansions. This implies a stronger acceleration of the wind (Figures~\ref{fig_profil_isopoly_fexp_example} (a,c), \ref{fig_u(r)_reajut_f-subsup_to_PSP_helios_data} (a)). The acceleration is also stronger if the flux tube expansion is present at larger distance because the enhanced pressure gradient overcomes more easily the weaker gravity (Figure~\ref{fig_profil_isopoly_fexp_example} (c)). 

Finally, the panel (d) of Figure~\ref{fig_u(r)_reajut_f-subsup_to_PSP_helios_data} shows that the median $f(r)$ profiles are not as smooth as the one determined in Figure~\ref{appendix:f(r)_profiles_details}. They are computed on a smaller subset (of $\sim 5 - 350$ profiles depending on the subset), which explains such weakly fluctuating shapes.  
However they still constitute a reliable representation of each v-f bin's typical expansion since no strong discrepancies between them are observed.

%%%%%%%%%%%%%%%%%%%%%%%%%%%%%%%%%%%%%%%%%%%%%%%%
\subsection{Implications of the updated isopoly profiles} 
\label{subsec_Implications_updated_isopoly_profiles}
%%%%%%%%%%%%%%%%%%%%%%%%%%%%%%%%%%%%%%%%%%%%%%%%

% Analysis of the speed profile decelerating region
The measurements from the interplanetary medium are closely modeled by the updated isopoly models as shown in Figure~\ref{fig_u(r)_reajut_f-subsup_to_PSP_helios_data}. Nevertheless, some of these updated models show below $\sim 7 \: \rs$ relatively high-speed regions (>250-500 km/s), and a local decrease in wind speed, so we may wonder how this could be consistent with remote-sensing observations of the solar corona. 

%It appears that 
In fact, several authors have observed a deceleration region in the wind radial evolution at distance lower than $\sim 7 \:\rs$ \citep{imamura2014, bemporad2017, casti2023}. These observations tend to support our modeling, in particular since they all focus on periods of rising solar activity, as does the present study. Moreover, they use different wind speed determination techniques. For instance \citet{imamura2014} use radio scintillation technique, while 
\citet{bemporad2017} and \citet{casti2023} apply the Doppler dimming technique. This observed deceleration region on such a radial interval ($\lesssim 7 \:\rs$) could illustrate both f-subsonic and f-supersonic deceleration regions, below $\rss$, and between $\rss$ and $7 \: \rs$ respectively. 

% Updated isopoly parameters 
The inclusion of the flux tube expansion $f(r)$ has modified the deduced coronal temperatures as summarized in Table \ref{tab_updated_isopoly_param}. Slow winds (\textbf{A} and \textbf{B} populations) do not see their respective proton coronal temperature $\Tpo$ vary significantly. However, $\Tpo$ significantly decreases for intermediate and fast winds (populations from \textbf{C} to \textbf{E}). More precisely, recalling that $\Tpo = [0.65, \: 1.10, \: 1.63, \: 2.51, \: 5.61]$ MK with $f(r)=1$ for the five wind populations studied in \citet{dakeyo2022}, the faster the wind, the more $\Tpo$ is reduced. The temperature decrease has a maximum of 3.4 MK for the wind \textbf{E}, leading to an isopoly fast wind with $\Tpo = 2.31$ MK. The latter temperature is more in accordance to the observed one in coronal holes. This supports the assumption that the coronal temperature may decrease from the f-subsonic to the f-supersonic solution, as presented for the parametric study in Sect.~\ref{subsec_fast_wind_large_f}.

% Mean proton temperature for isopoly model
For simplicity of computation, the present study has only investigated modified proton parameters while keeping electron parameters unchanged. And we recall that the isopoly parameters are not considering coronal constraints \citep{dakeyo2022}. Consequently to compare more realistically to coronal temperatures inferred from in-situ measurements of the heavy ion charge-state ratios, our coronal isopoly temperatures should be considered an equivalent mean temperature $\To$ at the base of the corona, considering $\To = (\Tpo + \Teo) /2$.
The five isopoly mean temperatures $\To$ are shown on the third line of the Table \ref{tab_updated_isopoly_param}. In particular, $\To$ of the fast wind (population \textbf{E}) can be as low as 1.6 MK while a velocity of 620 km/s is achieved at 1 au. As a future improvement, constraining the isopoly parameters by the coronal temperature derived from remote observations and charge state measurements could be an important step to reinforce the reliability of the isopoly model, and may lead to robustly constrained solar wind models.

%%%%%%%%%%%%%%%%%%%%%%%%%

%%%%%%%%%%%%%%%%%%%%%%%%%%%%%%%%%%%%%%%%%%%%%%%%%%%%%%%
\section{Discussion and Conclusion} \label{sec:conclusion}
%%%%%%%%%%%%%%%%%%%%%%%%%%%%%%%%%%%%%%%%%%%%%%%%%%%%%%%

%%%%%%%%%%%%%%%%%%%%%%%%

This paper presents a statistical analysis of the speed - expansion factor relation (v-f relation), using Solar Orbiter observations and ADAPT magnetograms during minimum solar activity. For this purpose we have established the magnetic connectivity from 01/08/2020 to 17/03/2022 when possible, using interplanetary streamline tracing and PFSS reconstruction. We assign a solar source origin to each 30-minute average measurement. In order to more realistically reproduce the wind streamline trajectories from the observations, we use isopoly modeling from \citet{dakeyo2022} to describe the radial evolution of wind properties and \citet{weber_davis1967} equations to model the tangential evolution. 

We find out that mixing all types of wind sources, there is only a weak global anti-correlation between the bulk speed and the expansion factor estimated at the source surface ($\rss =2.5~\rs$). Consequently, studies on v-f correlation conducted on coronal holes cannot be generalized to all types of wind sources.
The expansion factor on the source surface ($\fss$) should not be the main proxy parameter to infer the asymptotic wind speed and the radial wind acceleration profile. 

We have shown the existence of a fast solar wind population originating in high magnetic field regions mainly located at low latitudes, embedding a large expansion factor and lower densities than fast winds with low $\fss$. We explain the existence of such a wind with the generation of supersonic flows already in the low corona (below 2.5 $\rs$). The coronal super-radial expansion provides an important additional source of acceleration which drives supersonic wind below 2.5 $\rs$, resulting in a larger asymptotic velocity.   
The super-radial expansion in the supersonic regime can efficiently convert the thermal energy of the wind into kinetic energy. These type of solutions have been described as "f-supersonic" solutions, in opposition to "f-subsonic" fully subsonic below 2.5 $\rs$. 

We perform a parametric study based on the expansion profile described in \citet{kopp_holzer1976}. We show that the bulk speed at 1 au is monotonically increasing with larger maximum expansion value $f_m$, a shorter extension $\sige$ of the expansion region, and a larger radius $\re$ (marking almost the end of super expansion region).  
We conclude that the f-supersonic solutions require lower coronal temperature to reach the same asymptotic speed than the f-subsonic solutions. 

Next, we find the solar source associated to Solar Orbiter in-situ observations. We derive the $f(r)$ profiles from PFSS computations of the coronal magnetic field. Then, we incorporate these $f(r)$ profiles in the isopoly modeling. We find that fast wind is almost exclusively associated with f-supersonic solutions. We compute the expansion radius $\rexp$ defined as the radius where the expansion $f(r)$ increases the most. We derive that f-supersonic solutions have on average a higher $\rexp$ than f-subsonic ones, synonymous that f-supersonic wind types originate in average from sources surrounded by closed magnetic structures of higher altitude. This supports the importance of the expansion radius raised by our parametric study. 
Therefore, the f-supersonic modeling and the super radial typical expansion radius $\rexp$ constitute serious tracks to study the acceleration processes in the fast solar wind.  
The analysis of the uncertainties in the backmapping process, presented in Appendix~\ref{appendix:backmapping_uncertainties},  confirms our above conclusions.  

As another outcome, we find increasingly large initial bulk velocities $u_0$ for faster solar wind (Table $\ref{tab_updated_isopoly_param}$). For the faster wind they are in the range 40-100 km/s. Such high velocities are coherent with the scenario of magnetic interchange reconnection at the base of the wind as suggested by \citet{Gannouni2023, Bale_2023}. This is an interesting perspective to further investigate. In particular to quantify to what extent f-supersonic modeling and interchange reconnection mechanisms could work as complementary ingredients to provide an important acceleration low down in the corona, then to explain why the fast wind is reaching almost its terminal speed so close to the Sun.  
 
Finally, for the slow solar wind it is also important to take into account the $f(r)$ profile, derived from coronal field models, since part of the slow wind comes from narrow open-field corridor located at the border of active regions \citep[][and references therein]{Baker_2023}. In these regions, the presence of strong closed magnetic field shapes the $f(r)$ profile to get a strong expansion low down in the corona. Moreover, remote sensing velocity, temperature and density measurements are available to constraint the low coronal part of the models \citet{Cranmer1999spectroscopic, Cranmer2002coronal, imamura2014, bemporad2017, casti2023}. 
Then, we anticipate significant future progress in modeling both slow and fast winds with the isopoly model and taking into account the relevant coronal observational constraints.

\begin{acknowledgements}
    This research was funded by the European Research Council ERC SLOW\_SOURCE (DLV-819189) project. This research was supported by the International Space Science Institute (ISSI) in Bern, through ISSI International Team project \#463 (Exploring The Solar Wind In Regions Closer Than Ever Observed Before) led by L.~Harra. 
    This work was supported by CNRS Occitanie Ouest and LESIA. D.V.~is supported by STFC Consolidated Grant ST/W001004/1. This work made use of the Magnetic Connectivity Tool provided and maintained by the Solar-Terrestrial Observations and Modelling Service (STORMS). This work utilizes data produced collaboratively between AFRL/ADAPT and NSO/NISP. We recognize the collaborative and open nature of knowledge creation and dissemination, under the control of the academic community as expressed by Camille Noûs at http://www.cogitamus.fr/indexen.html. We thank the instrumental Solar Wind Analyser team (SWA) for valuable discussions. We acknowledge very valuable discussion with Miho Janvier, who has opened the problematic leading to this study.
\end{acknowledgements}


%\input{main.bbl}
%\bibliography

\bibliographystyle{aa}
%\typeout{}
%\bibliography{bibliography}

\appendix

%%%%%%%%%%%%%%%%%%%%%%%%%

%%%%%%%%%%%%%%%%%%%%%%%%%%%%%%%%%%%%%%%%%%%%%%%%%%%%%%%
\section{Backmapping uncertainties}
\label{appendix:backmapping_uncertainties}
%%%%%%%%%%%%%%%%%%%%%%%%%%%%%%%%%%%%%%%%%%%%%%%%%%%%%%%

% Uncertainties and artificial streamline deviation
The magnetic backmapping process is known to be sensitive to a number of effects that affect the propagation of the wind and can alter its speed, density and temperature radial evolution \citep{nolte_roelof1973, weber_davis1967, SanchezDiaz2016, macneil2022, dakeyo2024}. This results in a potential deviation of the mapped longitude in streamline computation. We have taken into account the influence of acceleration and corotational effects as detailed in Sects.~\ref{sec:isopoly_equations} and~\ref{sec:weber_davis_equations}. However, other effects such as stream-stream and ICME - stream interaction are more difficult to quantify in a statistical study. The uncertainties are also dependent on the speed of the modeled stream. With all these considerations, it may not be relevant to estimate a single global uncertainty value over the entire process of the statistical study, from the location of the probe until $\rs$. In order to provides uncertainties estimate related to our study, we present below different methods to estimate the reliability of the entire mapping process.

%%%%%%%%%%%%%%%%%%%%%%%%%%%%%%%%%%%%%%%%%%%%%%%%
\subsection{Mapping coordinates angular spread and  source identification}
\label{appendix:source_angular_spread_method}
%%%%%%%%%%%%%%%%%%%%%%%%%%%%%%%%%%%%%%%%%%%%%%%%

% Why looking to angular spread of the coordinates
The result of the magnetic connectivity process is sensitive to a coordinates displacement of a given field line at the source surface height. Indeed, %tracing back the field line path from the source surface down the Sun's surface, 
a coordinates displacement at $\rss$ may results in a significant difference in footpoints location at $\rs$.  
For this purpose, we aim to study the footpoints location variability at $\rs$ depending the fieldline coordinates displacement at $\rss$. To do this, we regroup the data points that are associated to the same source, and we study the spread of the source coordinates. 
This source extension is based on an angular step threshold applied on the coordinates.

% Source classification technique 
The source classification method is computed as follows.
The statistical backmapping study technique provides information on the time evolution of the mapped coordinates ($\thetao, \phio$) at the Sun's surface. The mapped coordinates displacement $(\dthetao, \dphio)$ are expected to evolve approximately as the angle scanned by the probe at $\rss$. Then, with continuous observations at a constant time cadence $\dt$, $(\dthetao, \dphio)$ are expected to evolve smoothly with time.
However, when connectivity changes from one source to another, we expect to see a jump of the footpoints coordinates, resulting in much larger values of $(\dthetao, \dphio)$ over the same time interval $\dt$. This jump in angular displacement marks the source boundaries in our source identification method. 

% Angular displacement and threshold criterion
Since a source can be angularly extended both in latitude and longitude, we define the angular distance~:
\begin{equation}
    \dang = \sqrt{\dthetao^2 + \dphio^2},
\end{equation}
equivalent to a norm of the angular displacement in both directions.
We set as a threshold $\dang > 10\degree$ as source change criteria. The data that are not associated to a source observed at least during 0.5 days, are not used in the following statistics. We assume that below this duration, the source is not enough broad to infer realistic source angular extension.

% Angular source expansion estimation
The total angular spread $\ars$ at $\rs$ is defined as the maximal angular distance covered for this given source :
\begin{align}
    \ars &= \sqrt{  (\text{max}(\thetao) - \text{min}(\thetao) )^2 
                 + ( \text{max}(\phio) -\text{min}(\phio) )^2 } \nonumber \\
        &= \sqrt{  (\text{range}(\thetao))^2 
                  + (\text{range}(\phio) )^2 }
\end{align}

The total angular spread $\arss$ is similarly defined at $\rss$. 
To analyze the change of the angular spread between $\rs$ and $\rss$, we define the angular spreading ratio :
\begin{align}
    \Qang = \frac{\arss}{\ars}
    = \sqrt{  \frac{ (\text{range}(\thetarss))^2 
                 +   (\text{range}(\phirss) )^2 }
                   { (\text{range}(\thetao) )^2 
                 +   (\text{range}(\phio) )^2 } 
           }
    \label{eq:Qang_definition}
\end{align}

% Meaning of the \Qang quantity 
For a given source, the quantity $\Qang$ is partly %not directly 
related to the mean expansion factor value since $\Qang$ is obtained by scanning the source extension, at both $\rs$ and $\rss$. 
Then, statistically, the field line centroid shift is expected to be greater (resp. smaller) for large (resp. small) $\fss$ values. 
$\Qang >1$ (resp. $<1$) for a source which is more (resp. less) extended at the source surface than at the photosphere (linking the two regions by field lines).

\begin{figure}[t!]
    \centering
    \hspace{-.3 cm}
    \includegraphics[width = 9cm]{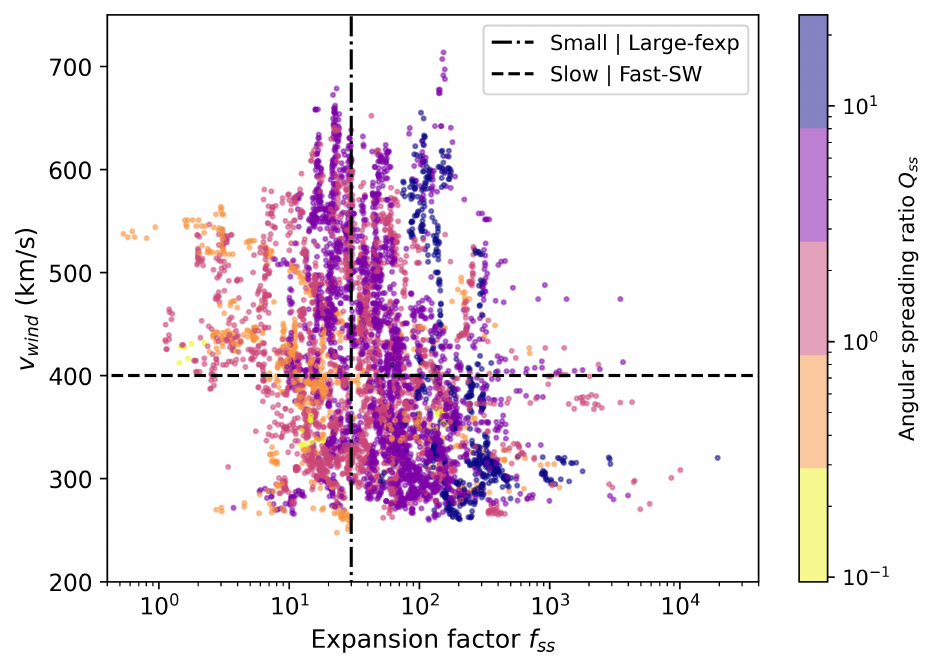}
    \caption{ Relationship between the measured velocity and the expansion factor, as in Fig.~\ref{fig_u_fexp_B0_magneto_low_high_latitude}, but colored by the angular spreading ratio $\Qang$  between $\rs$ and $\rss$ as defined by Eq.~\eqref{eq:Qang_definition}.  }   
    \label{fig_u_fexp_dang_compar_high_lat_large_source}
\end{figure}

\begin{figure*}[ht!]
    %\centering
    \hspace{0cm}
    \includegraphics[width = 18.5cm]{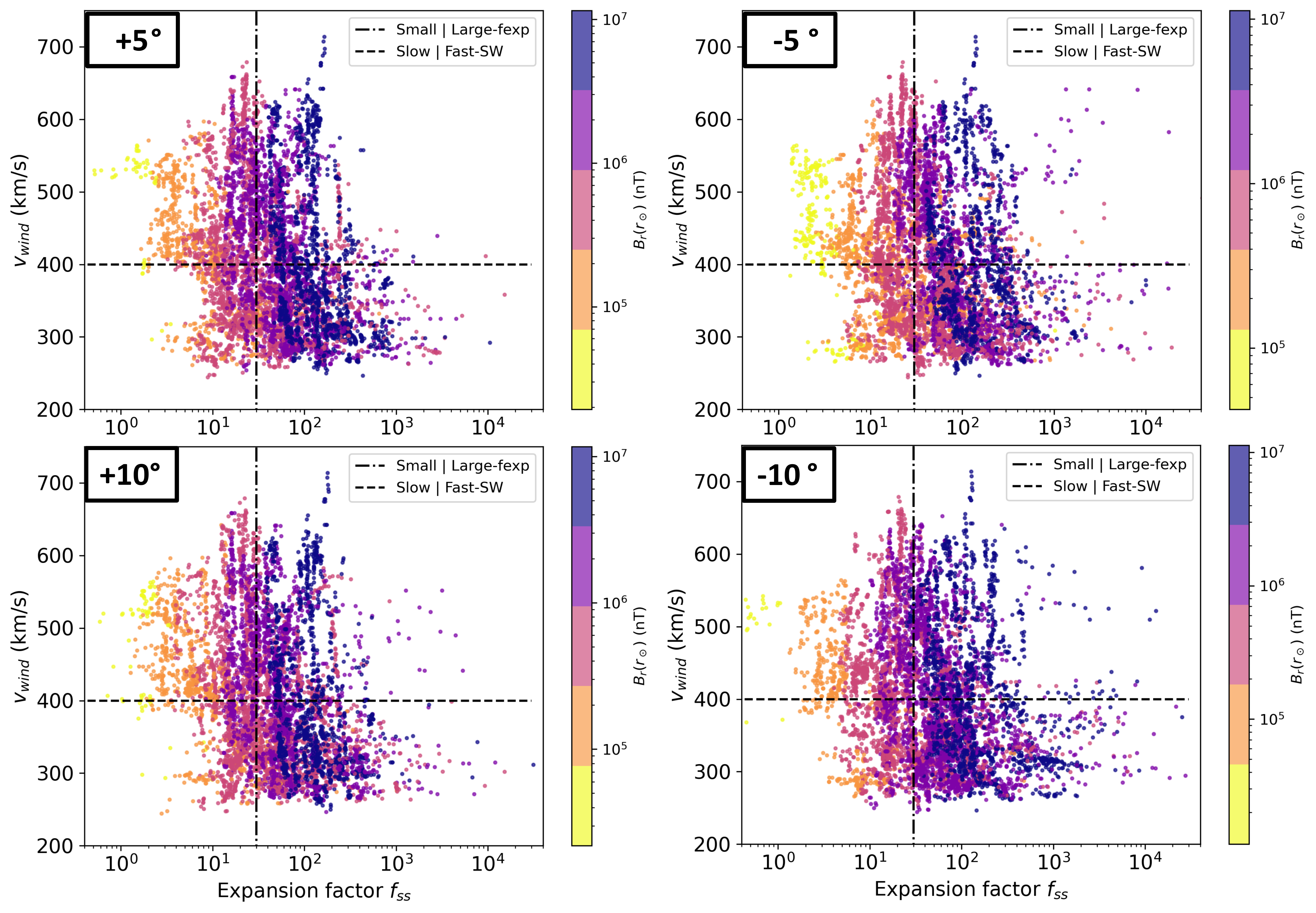}
    \caption{ Effect of a systematic bias on the mapping at the source surface. The same data than in panel (a) of Fig.~\ref{fig_u_fexp_B0_magneto_low_high_latitude} are used but with an artificial systematic deviation of $\pm$5\degree and $\pm$10\degree at $\rss$ on the field line tracing. }
\label{fig_u_fexp_B0_magneto_pm5_10}
\end{figure*}

% Angular spreading ratio to mappinf variability
The results are shown in the Figure ~\ref{fig_u_fexp_dang_compar_high_lat_large_source}.
 
The same data points than in Fig.~\ref{fig_u_fexp_B0_magneto_low_high_latitude} are shown except that $\sim$~13$\%$ of the mapped data are filtered out with since they are not belonging to a source having a duration of more than 0.5 day. 
As expected, $\Qang$ is correlated with $\fss$.

% Mapping variability and source angular spread
Moreover, considering a 1 degree angular extension at $\rs$, the value of $\Qang$ is a direct measure of the angular extension at $\rss$ in degrees of the source coordinates. Consequently, the larger $\Qang$, the less mapped coordinates $(\thetao,\phio)$ are sensitive to uncertainties of the mapping process, since there is a focusing effect of the field line mapping from the source surface to the photosphere. Then, sources with large $\Qang$ are localized with greater accuracy throughout the entire backmapping process (from the spacecraft to $\rs$).

%%%%%%%%%%%%%%%%%%%%%%%%%%%%%%%%%%%%%
\subsection{Unquantified streamline deviations}
\label{appendix:backmapping_uncertainties_unquantified_devia}
%%%%%%%%%%%%%%%%%%%%%%%%%%%%%%%%%%%%%

We estimate below to what extent streamline deviation from unquantified effects would affect our final results. Consequently we have re-computed the backmapping study and results from Sect.~\ref{sec:Results} with an artificial deviation of the longitude $\phi(\rss)$ of $\pm$5\degree and $\pm$10\degree. They represent an hypothetical systematic bias. All the other parts of the magnetic connectivity process are the same as presented in Sect.~\ref{sec:connectivity_methods}. 
We observe in Fig.~\ref{fig_u_fexp_B0_magneto_pm5_10} that the global shape presented in Fig.~\ref{fig_u_fexp_B0_magneto_low_high_latitude} is similar for the $\pm$5\degree and $\pm$10\degree panels. Some variability is mostly present for large $\fss$. %and high speed, 
However, this concerns few mapped data. We also notice that the low $\vwind$ and low $\fss$ region is less filled in the +5° and +10° maps, while more filled in for the -5° and -10° ones. 

The large $\vwind$ and large $\fss$ region is still present for all four cases with similar $\Brs$ and $\fss$ values. Consequently, when considering relatively large variability in the mapping process, the fast wind associated with large $\fss$ regions is still observed. This statement is also supported by the fact that they typically have a large $\Qang$ (defined by Eq.~\eqref{eq:Qang_definition}), as shown in Figure ~\ref{fig_u_fexp_dang_compar_high_lat_large_source}, so that the variability in streamline tracing, from the spacecraft to $\rss$, poorly affect the location of the corresponding footpoint at $\rs$.

\begin{figure*}[ht!]
    \hspace{-.4cm}
    \includegraphics[width = 18.5cm]{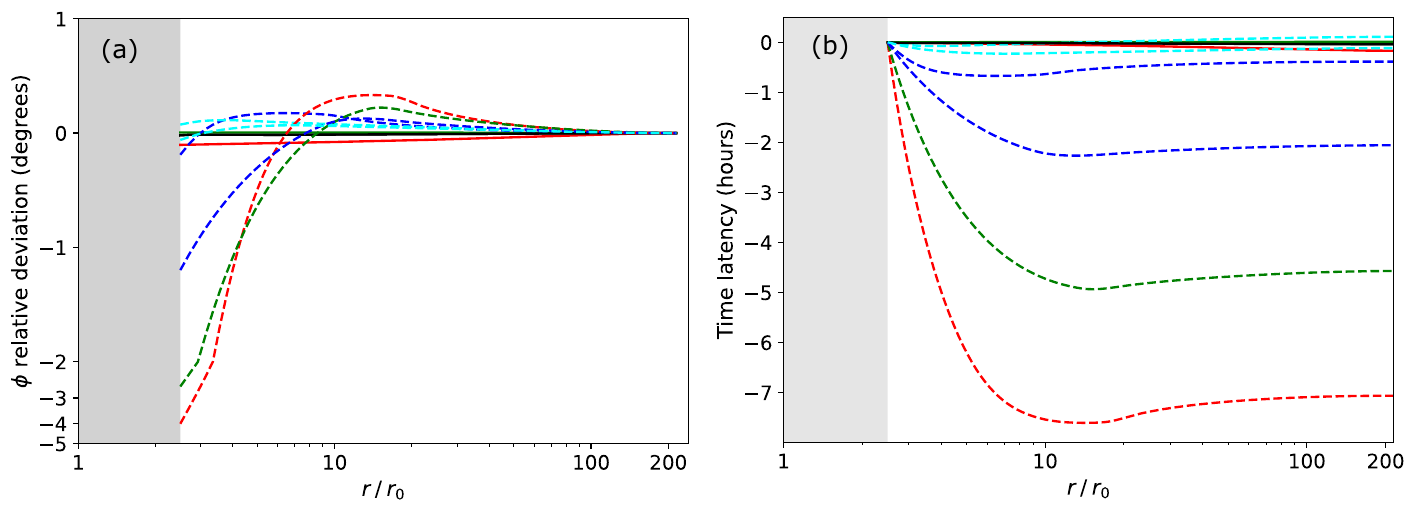}
    \caption{Streamline and travel time discrepancies between the 5 isopoly wind profiles used in the computation of the backmapping process \citep{dakeyo2022}, and the updated 5 isopoly profiles accounting for super radial expansion presented in Figure~\ref{fig_u(r)_reajut_f-subsup_to_PSP_helios_data}. Panel (a) : Streamline angle deviation from 1 au to $\rss$; Panel (b) : Travel time latency. 
    The f-subsonic and the f-supersonic solutions are plotted in solid and dashed lines respectively.  The wind populations, from the slowest to the fastest are shown with black, red, green, dark blue and light blue.
    }
\label{fig_devi_streamline_fsub_super}
\end{figure*}

%%%%%%%%%%%%%%%%%%%%%%%%%

%%%%%%%%%%%%%%%%%%%%%%%%%%%%%%%%%%%%%%%%%%%%%%%%%%%%%%%
\subsection{Backmapping uncertainties : f-supersonic and f-subsonic deviations}
\label{appendix:backmapping_uncertainties_f_super_f_sub}
%%%%%%%%%%%%%%%%%%%%%%%%%%%%%%%%%%%%%%%%%%%%%%%%%%%%%%%

%%%%%%%%%%%%%%%%%%%%%%%%

% Longitude misestimation from f-sub f-super
The change of speed profile from f-subsonic to f-supersonic affects both the streamline calculation and the travel time, and so the mapping results. To estimate the resulting longitude deviation and time latency between f-subsonic and f-supersonic solutions, we compute the streamlines of the updated isopoly profiles presented in Fig.~\ref{fig_u(r)_reajut_f-subsup_to_PSP_helios_data}, and the travel time latency between the two types of solutions. We compare the updated streamlines and travel times with the initial ones of the isopoly profiles, used for the connectivity calculation of \citet{dakeyo2022}. This provides a post-mapping estimate of a possible supplementary error.  

In order to compare the updated isopoly profiles with those of \citet{dakeyo2022}, since the asymptotic speed of each profile of the same wind population is very close to, but not exactly similar to, that of \citet{dakeyo2022}, we normalize each updated profile by the 1 au speed of its reference profile of \citet{dakeyo2022}. 
Regarding that the normalization coefficients are very close to 1 ($\leq 1.08$ i.e. small readjustment), we assume that it does not significantly affect the shape of the normalized profiles, which remain consistent with the non-normalized wind population speed evolution. This allows each wind population to have the same 1 au speed and set a longitude deviation from the same reference.

Panel (a) of the Figure~\ref{fig_devi_streamline_fsub_super} presents the longitude deviation, and panel (b) shows the time latency.  They are computed as the difference of results between the updated isopoly models and the ones computed by \citet{dakeyo2022}. The colors code is the same as the one in the Figure~\ref{fig_u(r)_reajut_f-subsup_to_PSP_helios_data}. 

%A deviation of zero degrees assumes no deviation on the streamline tracing. We observe 
The slowest and fastest wind, populations \textbf{A} and \textbf{E} respectively,  present deviations much lower than 1\degree. These lasts are lower than the 1\degree resolution of the used magnetograms, then such deviations are imperceptible in the mapping results. 
Next, we find a streamline deviation of about 3-4\degree, at $\rss$ for the wind populations \textbf{B} and \textbf{C}. Indeed, in Figure~\ref{fig_u(r)_reajut_f-subsup_to_PSP_helios_data}, populations \textbf{B} and \textbf{C} have the largest difference of the velocity profile between f-sub and f-supersonic solutions. % and \textbf{D}.
% Regarding the wind populations with larger deviations ($\leq$ 4 \degree), 
However, the uncertainties study presented in Appendix~\ref{appendix:backmapping_uncertainties_unquantified_devia} shows that a systematic deviation lower than 5\degree\ is not changing the overall v-f relation results. 

The panel (b) of the Figure~\ref{fig_devi_streamline_fsub_super} indicates that all updated f-subsonic solutions including $f(r)$ modeling do not present any travel time latency with reference the isopoly model of \citet{dakeyo2022}. 
However, the f-supersonic profiles shows small to intermediate time latency, from $\sim$ 2 hours for fast wind, until 5 to 8 hours for intermediate and slow solar wind. This implies that a different magnetogram should a priori be used to compute the coronal magnetic field.  
However, the consecutive magnetograms taken every 6 hours, do not present in general strong changes, in particular during low solar activity, as analyzed here. 
Moreover, this 5 to 8 hours time latency only concerns a relatively small part of the observed wind (f-supersonic solutions of winds \textbf{B} and \textbf{C}). Thus, it is expected to result in a small shift of part of the f-supersonic solutions for the \textbf{B} and \textbf{C} 
wind populations, but with almost no effect for other wind populations. 
This would introduce only small and localized changes on the v-f relation presented in  Figs.~\ref{fig_u_fexp_B0_magneto_low_high_latitude} and~\ref{fig_u_fexp_n_r2}. %present article.

Next, we notice that for the f-subsonic solutions, the modification of the speed profile exclusively takes place below $\rss$. Then, including $f(r)$ profile in the wind model affects neither the spiral nor the travel time. 
% No influence on streamline f-sub f-super
Finally,  we conclude that the mapping deviation due to a change from f-subsonic to f-supersonic modeling is negligible on the statistical results presented in this article.

%%%%%%%%%%%%%%%%%%%%%%%%%

%%%%%%%%%%%%%%%%%%%%%%%%%%%%%%%%%%%%%%%%%%%%%%%%%%%%%%%
\section{PFSS expansion factor profile}
\label{appendix:f(r)_profiles_details}
%%%%%%%%%%%%%%%%%%%%%%%%%%%%%%%%%%%%%%%%%%%%%%%%%%%%%%%

\begin{figure*}[t]
    \centering
    \hspace{-.5cm}
    \includegraphics[width = 18.3cm]{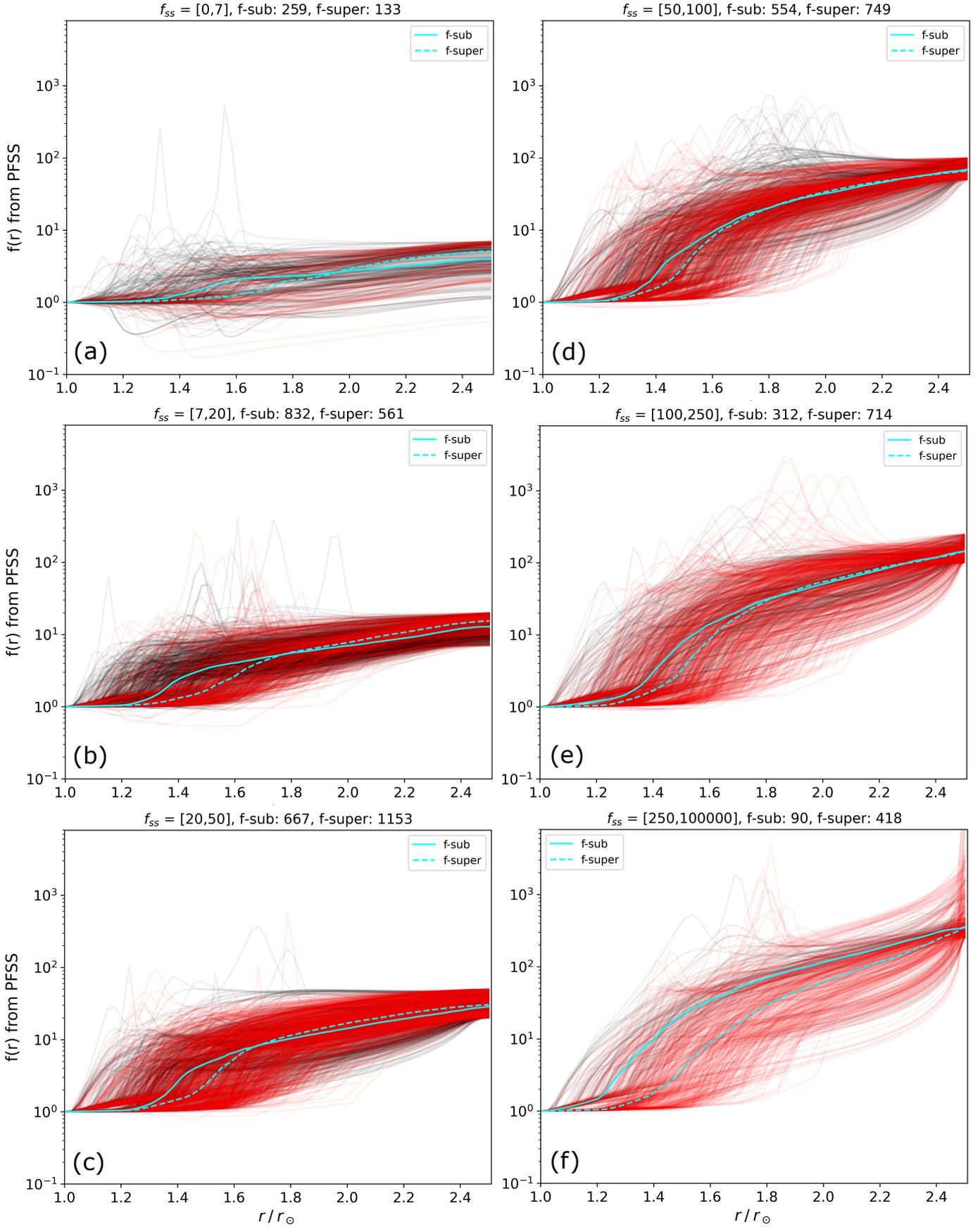}
    \caption{Expansion factor profiles $f(r)$ and their median profiles, based on f-subsonic and f-supersonic classification from the panel (a) of Figure~\ref{fig_u_fexp_obs_supersonic_hexp}. The $f(r)$ profiles associated with f-subsonic and f-supersonic solutions are displayed in black and red, and their respective median $f(r)$ profiles are plotted in cyan solid line and cyan dashed line, respectively. The number of profiles used to compute the median profiles is displayed on the top of each panel. 
    The panel (a) to (f) correspond to $\fss$ interval with $\fss$ bins limits $[7, 20, 50, 100, 250]$ respectively, setting 6 $\fss$ intervals.
    The larger boundary of panel (f) is set to a considerably larger value to represent infinity.
    }
    \label{fig_fprof_median_of_vf_plot_from_PFSS_fbins}
\end{figure*}

The $f(r)$ profiles, computed with the PFSS extrapolation of magnetograms, have a broad variety of shapes. Since the results with the analytical profile of \citet{kopp_holzer1976} in Sect.~\ref{subsec_fast_wind_large_f} show that the maximum expansion parameter is one of the key parameter, but not the only one, we divide the results into several $\fss$ bins, to study the effects of the other expansion parameters. We split the data with the $\fss$ values 7, 20, 50, 100, and  250. Figure~\ref{fig_fprof_median_of_vf_plot_from_PFSS_fbins} shows all $f(r)$ profiles separated in 6 categories. As mentioned in Sect.~\ref{subsec_fast_wind_large_f}, for each f(r) profile an isopoly model is computed (as many models as profiles) and then assigned to f-subsonic and f-supersonic solutions. 
The $f(r)$ profiles associated with f-subsonic and f-supersonic solution are plotted in black and red solid lines, respectively (as in panel (a) of Figure 
\ref{fig_u_fexp_obs_supersonic_hexp}). 

A wide variety of profiles are present, although they are only partially visible due to the superposition of many cases. Then, we compute the median profiles, calculating the median value at each radial distance, of a given $\fss$ bin, either between all the f-subsonic or f-supersonic f(r) profiles.  
This allows to derive the most typical expansion profiles, given a $\fss$ bin, and a type of isopoly solution. They are plotted in cyan solid and dotted lines for f-subsonic and f-supersonic, respectively.
All these median profiles have a similar shape with a monotonous behavior. They all are nearly flat for small $r$ values ($r<1.3\,\rs$), then $f(r)$ increases rapidly with $r$, to increases more steadily with $r$ at larger $r$ values up to $\rss$.  Then, all these monotonous profiles are mostly characterized by $f(\rss)$ and by the radius $\rexp$ with the largest slope (defined by Eq.~\eqref{eq_def_rexp}). The main difference is a larger $\rexp$ for f-supersonic solutions by about $0.2~\rs$ compared to f-subsonic ones. 
    
% $f(r)$ profiles description
Analyzing the individual profiles, some of the $f(r)$ reconstructed from PFSS have non-monotonic profiles, with initially diverging then reconverging flux tube (visible in all panels), and the opposite case with a converging then diverging flux tube (mostly panels (a) and (b)). 
For example, the complexity of $f(r)$ across a corridor of open flux is shown in Fig. 4 of \citet{Baker_2023}. The presence of closed fields of different magnetic flux and spatial extension around an open field region implies a differential expansion of the open flux. 
Some profiles have a more complex shape with possible multiple bumps of $f(r)$ profiles. These non-monotonic profiles could be associated with closed magnetic structures surrounding the open field line, exerting a magnetic pressure that could pinch the flux tube. Several magnetic structures surrounding the same open magnetic field line may create several bumps on the flux tube expansion, so on the $f(r)$ profile.

\end{document}